\documentclass[twocolumn,amsmath,amssymb,floatfix,superscriptaddress,prl]{revtex4-2}

\usepackage{bm,graphicx,url,epsf,color}
\usepackage{amssymb}
\usepackage{amsthm}
\usepackage{cases}
\usepackage[colorlinks=true,linkcolor=blue,citecolor=blue]{hyperref}
\usepackage{comment}

\begin{document}
\title{Many-body quantum interference route to the two-channel Kondo effect:\\Inverse design for molecular junctions and quantum dot devices}


\author{Sudeshna Sen}
\affiliation{Department of Physics, IIT(ISM) Dhanbad, Dhanbad-826004, Jharkhand, India}
\author{Andrew K. Mitchell}\email[]{Andrew.Mitchell@UCD.ie}
\affiliation{School of Physics, University College Dublin, Belfield, Dublin 4, Ireland}
\affiliation{Centre for Quantum Engineering, Science, and Technology, University College Dublin, Ireland}


\begin{abstract}
\noindent Molecular junctions -- whether actual single molecules in nanowire break junctions or artificial molecules realized in coupled quantum dot devices -- offer unique functionality due to their orbital complexity, strong electron interactions, gate control, and many-body effects from hybridization with the external electronic circuit. Inverse design involves finding candidate structures that perform a desired function optimally. Here we develop an inverse design strategy for generalized quantum impurity models describing molecular junctions, and as an example, use it to demonstrate that many-body quantum interference can be leveraged to realize the two-channel Kondo critical point in  simple 4- or 5-site molecular moieties. We show that remarkably high Kondo temperatures can be achieved, meaning that entropy and transport signatures should be experimentally accessible. 
\end{abstract}
\maketitle


Nanoelectronics circuits are quantum devices featuring a nanostructure with a few confined and typically strongly correlated degrees of freedom coupled to source and drain metallic leads \cite{kouwenhoven2001few,hanson2007spins,goldhaber1998kondo,*cronenwett1998tunable,park2002coulomb,*liang2002kondo,perrin2015single}. For molecular junctions, a single molecule can bridge the gap in a nanowire \cite{champagne2005mechanically}. The electrical conductance of such a junction is controlled by the structure and chemistry of the molecule, through which a current must pass \cite{xu2003measurement}. A range of physics can be realized in such systems -- including Coulomb blockade \cite{thijssen2008charge} and various Kondo effects \cite{park2002coulomb,*liang2002kondo,paaske2006non,roch2009observation,parks2010mechanical,guo2021evolution,zalom2019magnetically,lucignano2009kondo,requist2014kondo}, quantum interference \cite{stafford2007quantum,guedon2012observation,li2019gate,bai2019anti,greenwald2021highly,mitchell2017kondo,pal2019nonmagnetic,li2019symmetry}, and phase transitions \cite{roch2008quantum,florens2011universal}. This presents the tantalizing possibility of devices at the limit of miniaturization that leverage inherently quantum effects to provide enhanced functionality as switches \cite{lortscher2006reversible,quek2009mechanically,van2010charge,zhang2015towards}, transistors \cite{park2002coulomb,*liang2002kondo,perrin2015single}, diodes and rectifiers \cite{elbing2005single,perrin2016gate,batra2013tuning,diez2009rectification,metzger2003unimolecular,gupta2023nanoscale}, and even as tools for chemical analysis \cite{dief2023advances}. A grand challenge is to find molecular species that can form robust junctions to perform a desired function optimally \cite{xin2019concepts}.

Simple artificial molecular junctions can also be fabricated in semiconductor coupled quantum dot (QD) devices \cite{kouwenhoven1995coupled,jeong2001kondo}. The design of such systems need not obey chemical structure principles \cite{feynman2018there}, and they benefit from \textit{in-situ} tunability \cite{barthelemy2013quantum,petropoulos2024nanoscale}. They can also be integrated with other components to realize more exotic effects, such as fractionalization at the two-channel Kondo (2CK) quantum critical point \cite{affleck1991universal}, which results from the frustration of screening when a single spin-$\tfrac{1}{2}$ degree of freedom is coupled to two independent conduction electron channels \cite{nozieres1980kondo}. The 2CK effect has gained prominence recently as a route to engineer many-body Majorana zero modes in nanostructures \cite{emery1992mapping,komijani2020isolating,lotem2022manipulating,lopes2020anyons}. Spectacular experimental realizations of 2CK physics in QD systems \cite{potok2007observation,keller2015universal,iftikhar2015two} have however required the use of a `quantum box' or metallic island to provide a reservoir of many interacting electrons \cite{Yuval_2003,furusaki1995theory}. Can the 2CK effect be realized in simpler QD systems without the use of these components? If so, what is the minimum number of interacting sites needed? Can we find molecular moieties that realize 2CK physics when placed in a junction?


\textit{Model.--} Molecular junctions and QDs are described by generalized quantum impurity models \cite{Hewson} of the form $\hat{H}=\hat{H}_{\rm mol}+\hat{H}_{\rm leads}+\hat{H}_{\rm hyb} + \hat{H}_{\rm gate}$. Here we formulate the isolated molecule as an extended Hubbard Hamiltonian, 
\begin{equation}\label{eq:Hmol}
    \hat{H}_{\rm mol} = \sum_{\sigma=\uparrow,\downarrow}\:\sum_{m,n} t_{mn}^{\phantom{\dagger}} d_{m\sigma}^{\dagger}d_{n\sigma}^{\phantom{\dagger}} + \tfrac{1}{2}\sum_{m,n} U_{mn} \hat{n}_m\hat{n}_n
\end{equation}
where $d_{m\sigma}^{(\dagger)}$ annihilates (creates) an electron on molecule orbital $m$ with spin $\sigma$ and $\hat{n}_m=\sum_{\sigma}d_{m\sigma}^{\dagger}d_{m\sigma}^{\phantom{\dagger}}$ is a number operator. Single-particle processes are parameterized by $t_{mn}$ whereas $U_{mn}$ embodies electronic interactions.
The gate voltage $V_g$ controls the charge on the molecule via $\hat{H}_{\rm gate}=V_g\sum_m \hat{n}_m$. The leads are described by continua of free fermions, $\hat{H}_{\rm leads}=\sum_{\alpha,\sigma} \epsilon_k^{\phantom{\dagger}} c_{\alpha\sigma k}^{\dagger}c_{\alpha\sigma k}^{\phantom{\dagger}}$ with $\alpha=s,d$ for source and drain. The molecule frontier orbital $d_{r_{\alpha} \sigma}$ couples to a local orbital $c_{\alpha\sigma}$ of lead $\alpha$ via $\hat{H}_{\rm hyb}=\sum_{\alpha,\sigma}V_{\alpha}(d_{r_{\alpha}\sigma}^{\dagger}c_{\alpha\sigma}^{\phantom{\dagger}}+{\rm H.c.})$, where $c_{\alpha\sigma}=\tfrac{1}{V_{\alpha}}\sum_{k}V_k c_{\alpha \sigma k}$. 

Strong electron interactions
\cite{goldhaber1998kondo,*cronenwett1998tunable,park2002coulomb,*liang2002kondo} produce rich many-body physics but also preclude brute force solutions \cite{Hewson}. Inverse design is challenging because physical properties then depend in a highly nontrivial way on the delicate interplay of many microscopic parameters. It is a formidable task to find a set of model parameters that yield specific device functionalities. However, if only the \emph{low-temperature} behavior is of interest, then simpler low-energy effective models may be used \cite{schrieffer1966relation,*bravyi2011schrieffer,rigo2020machine,rigo2024unsupervised}. The connection between effective model parameters and low-temperature physical properties is more transparent.

Here we focus on one such scenario, where the low-temperature physics that we seek is that of the 2CK critical point \cite{affleck1993exact}. The condition for obtaining this behavior in molecular junctions is simply stated in terms of the low-energy effective 2CK model. Inverse design then consists of finding the set of microscopic model parameters satisfying this condition. We show that this is achievable in remarkably simple systems, with just a few interacting degrees of freedom, and without the interacting electron reservoirs used previously in experiments \cite{potok2007observation,keller2015universal,iftikhar2015two}.


\emph{Effective models.--} An odd number of electrons can be accommodated on the molecule by tuning gate voltages, such that the ground state of $\hat{H}_{\rm mol}$ is a unique spin-doublet state. At low temperatures, effective spin-flip Kondo exchange interactions and potential scattering are generated, described by a generalized 2CK model \cite{SI},
\begin{equation}\label{eq:2ck}
    \hat{H}_{\rm eff} = \hat{H}_{\rm leads} + \sum_{\alpha,\beta} \left [J_{\alpha\beta\;}\hat{\vec{\mathbb{S}}}\cdot \hat{\vec{s}}_{\alpha\beta} + W_{\alpha\beta}^{\phantom{\dagger}}\sum_{\sigma} c_{\beta \sigma}^{\dagger} c_{\alpha \sigma}^{\phantom{\dagger}} \right ] 
\end{equation}
where $\hat{\vec{\mathbb{S}}}$ is a spin-$\tfrac{1}{2}$ operator for the molecule ground state doublet and  $\hat{\vec{s}}_{\alpha\beta} = \tfrac{1}{2} \sum_{ss'} c_{\beta s'}^{\dagger} \vec{\boldsymbol{\sigma}}_{s's} c_{\alpha s}^{\phantom{\dagger}}$ are conduction electron spin operators. We refer to the $J_{\alpha\beta}$ and $W_{\alpha\beta}$ terms as exchange and potential scattering, respectively. The form of Eq.~\ref{eq:2ck} is guaranteed by $SU(2)$ spin symmetry if only the most RG relevant terms are considered \cite{mitchell2017kondo}. Since $J_{sd}=J_{ds}$ and $W_{sd}=W_{ds}$ by hermiticity, the low-energy behavior of such molecular junctions is controlled by just six effective parameters. 

The 2CK critical point arises for equal antiferromagnetic Kondo interactions $J_{ss}=J_{dd}>0$, but when the source-drain mixing terms vanish, $J_{sd}=W_{sd}=0$ \cite{affleck1993exact,Sela_2011,*Mitchell_2012a}. In molecular junctions or coupled QD devices, the 2CK effect should be realizable when the molecule or QD has a net spin-$\tfrac{1}{2}$ ground state and when the effective model parameters satisfy these conditions. $W_{ss}$ and $W_{dd}$ are RG irrelevant and play no role in the following. 

\begin{figure}[t!]
\includegraphics[width=1.0\linewidth]{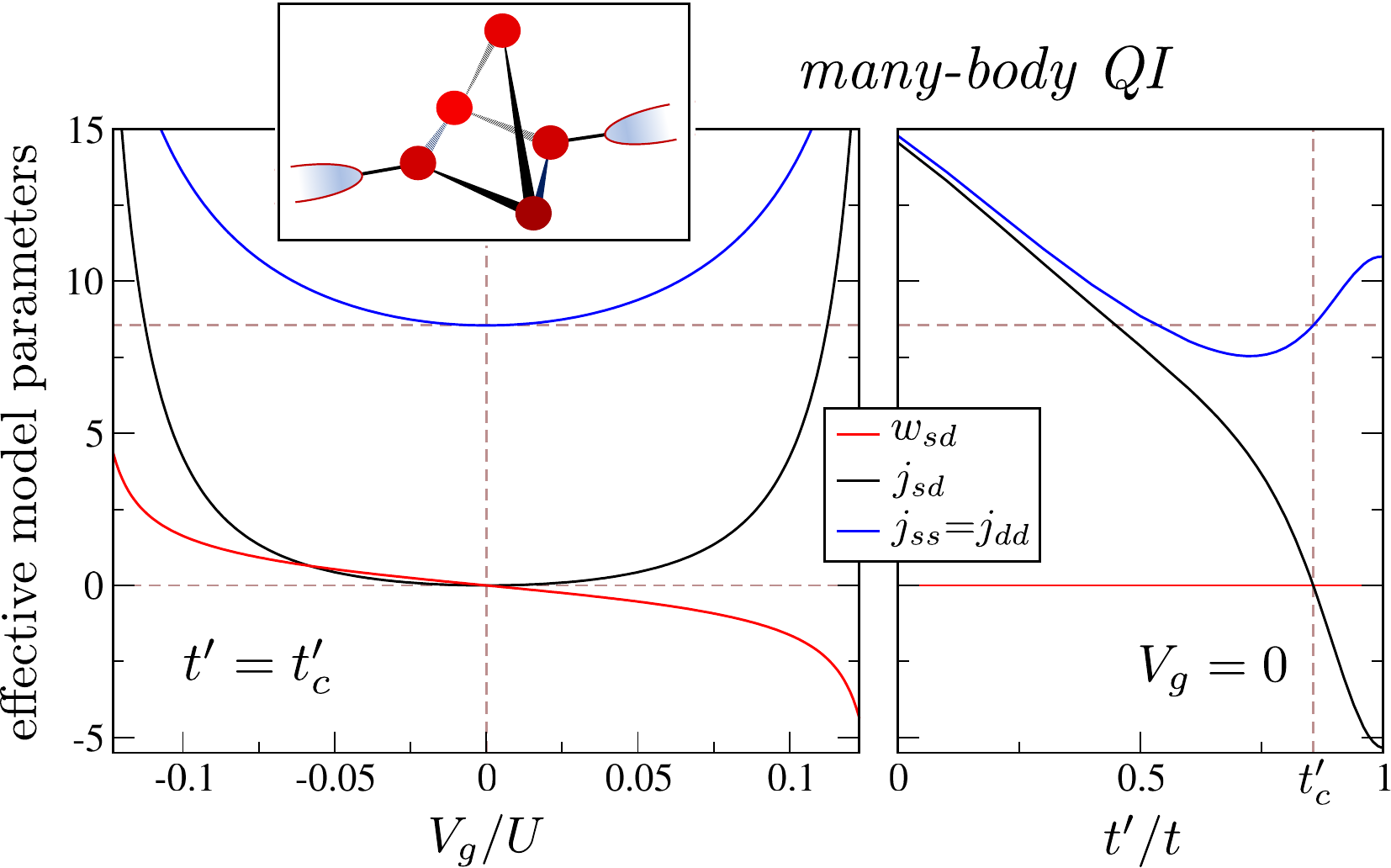}
  \caption{The simplest molecular moiety to exhibit the 2CK effect with 5 interacting active orbitals. The effective molecule-lead Kondo interactions $j_{ss}$ and $j_{dd}$ are equal and antiferromagnetic (blue line), while source-drain mixing terms vanish due to many-body QI. Potential scattering $w_{sd}$ (red) vanishes at gate voltage $V_g=0$ by particle-hole symmetry, whereas exchange cotunneling $j_{sd}$ (black) vanishes on tuning the couplings $t'/t$. Obtained here via SWT and plotted for $U/t=1$.}
   \label{fig:swt}
\end{figure}


\textit{Quantum interference (QI) and conductance nodes.--} 
Single-molecule junctions often exhibit QI phenomena, with the most dramatic effect being electrical conductance nodes due to the destructive interference of competing transport pathways through the molecule \cite{guedon2012observation,li2019gate,bai2019anti,greenwald2021highly}. However, such a description of the QI and transport is typically on the single-particle level encoded by the real-space hopping matrix $t_{nm}$ \cite{lambert2015basic}, and is inapplicable for interacting systems displaying Coulomb blockade or Kondo effects. 
Although sequential single-particle tunneling processes may be afflicted by decoherence at weak coupling \cite{thijssen2008charge}, coherent many-body processes are more robust at low temperatures in interacting systems \cite{konig2001coherence}.
Many-body QI \cite{pedersen2014quantum,mitchell2017kondo} is naturally richer than its single-particle counterpart, being defined in a high-dimensional Fock space, and provides new channels for QI (e.g.~between particles and holes). Many-body QI can cause any of the parameters $J_{\alpha\beta}$ and $W_{\alpha\beta}$ to vanish in the effective 2CK model Eq.~\ref{eq:2ck}. $J_{sd}=W_{sd}=0$ must produce a conductance node because then the charge in the leads is separately conserved. The 2CK critical point therefore arises at a conductance node, which can be driven by many-body QI. We dub this the QI-2CK effect.


\textit{Perturbative solution.--} 
We consider first the perturbative derivation of the effective 2CK parameters from those of the bare model by means of a generalized Schrieffer-Wolff transformation (SWT) \cite{schrieffer1966relation,*bravyi2011schrieffer}. This is done by projecting the full model for the junction onto the spin-doublet molecule ground states by eliminating virtual excitations to second order in $\hat{H}_{\rm hyb}$. In the Supplementary Material (SM) \cite{SI} we formulate this problem in an efficient way that does not require full diagonalization of $\hat{H}_{\rm mol}$, but only uses information on the ground state energy and wavefunction of the isolated molecule. Comparatively large systems can then be treated by using methods that target ground state properties \cite{gagliano1986correlation,schollwock2011density,zgid2012truncated}.

Eq.~\ref{eq:2ck} is obtained by SWT with effective parameters $J_{\alpha\beta}\equiv V_{\alpha}V_{\beta} j_{\alpha\beta}$ and $W_{\alpha\beta}\equiv V_{\alpha}V_{\beta} w_{\alpha\beta}$ that can be calculated from many-body scattering amplitudes $A^{\sigma\alpha\beta}=p^{\sigma\alpha\beta}-h^{\sigma\alpha\beta}$ which involve the tunneling of both particles ($p$) and holes ($h$) with spin-$\sigma$ through the molecule from lead $\alpha$ to lead $\beta$. We may write $j_{\alpha\beta}=2(A^{\uparrow\alpha\beta}-A^{\downarrow\alpha\beta})$ and $w_{\alpha\beta}=\tfrac{1}{2}(A^{\uparrow\alpha\beta}+A^{\downarrow\alpha\beta})$, with the $p$ and $h$ amplitudes obtainable in closed form as detailed in the SM \cite{SI}.

Many body QI can appear here in different ways: through the vanishing of individual $p$ or $h$ processes due to interference of competing Fock space propagators, by a cancellation of terms with different spin, or by a cancellation of $p$ and $h$ amplitudes for a given process. 

In fact, particle-hole ($ph$) symmetry guarantees the latter, since then  $p^{\sigma\alpha\beta}=h^{-\sigma\alpha\beta}$ and hence $W_{ss}=W_{dd}=W_{sd}=0$ in Eq.~\ref{eq:2ck}. A system is $ph$-symmetric when its Hamiltonian is invariant to the $ph$ transformations $d_{n\sigma}\to e^{i\phi_{n\sigma}} d_{n\sigma}^{\dagger}$ for all $n\sigma$ (with suitable phases $\phi_{n\sigma}$). The celebrated Coulson-Rushbrooke pairing theorem \cite{coulson1940note} is a statement about $ph$ symmetry, with $p$ and $h$ excitations appearing symmetrically around the ground state for molecules satisfying the `starring rule' \cite{tsuji2018quantum,pedersen2014quantum}. A system may exhibit $ph$ symmetry 
if the molecular structure encoded by the single-particle adjacency matrix $t_{mn}$ can be accommodated on a bipartite graph. Therefore $ph$-symmetric systems must \textit{not} have odd loops.


\emph{Satisfying the 2CK condition.--}
Since $ph$ symmetry implies $W_{sd}=0$, we search for $ph$-symmetric systems in which $J_{sd}=0$ can also be achieved.  In addition we want $J_{ss}=J_{dd}$ for the 2CK effect so we consider only $sd$-symmetric molecular moieties. As a simple starting point we study $M$-site Hubbard chains with constant nearest neighbour hopping $t$, local Coulomb repulsion $U$, and local potential $\epsilon=-U/2$. Leads $s$ and $d$ are connected to molecule sites $1$ and $M$. For odd $M$ the ground state around $V_g=0$ is a unique spin-doublet and we numerically perform the SWT as shown in the SM \cite{SI}. The system is $ph$-symmetric at $V_g=0$ such that $W_{\alpha\beta}=0$. We also find $J_{ss}=J_{dd}>0$. Although $J_{sd}$ is always finite, we find that its sign alternates for $M=1,3,5,7,...$. In particular, $J_{sd}<0$ for $M=3$ but $J_{sd}>0$ for $M=5$. One might anticipate that interpolating between $M=3$ and $M=5$ might yield a sweet spot solution where $J_{sd}=0$. Avoiding odd loops and preserving $sd$ symmetry, this can be achieved by connecting sites $1$ to $4$ and $2$ to $5$, viz:
\begin{align}
\label{eq:5site}
    H_{\rm mol} = \frac{U}{2} & \sum_{m=1}^5 \left (\hat{n}_m-1\right )^2  +t \sum_{\sigma}  \sum_{m=1}^4  \left( d_{m\sigma}^{\dagger} \nonumber d_{m+1\sigma}^{\phantom{\dagger}}+{\rm H.c.} \right )  \\
   + t' &\sum_{\sigma}\left( d_{1\sigma}^{\dagger} d_{4\sigma}^{\phantom{\dagger}}+ d_{2\sigma}^{\dagger} d_{5\sigma}^{\phantom{\dagger}} +{\rm H.c.} \right ) \;.
\end{align}
For small $t'/t$ we expect small perturbations to the $M=5$ chain solution, whereas for large $t'/t$ the next-next-nearest-neighbour tunneling provides a shortcut through the chain so that only $3$ sites are needed to connect $s$ and $d$ leads. Numerical results of the SWT are presented in Fig.~\ref{fig:swt}, together with a schematic illustration of the junction. At $ph$ symmetry $V_g=0$, the effective model parameters are plotted as a function of $t'/t$ in the right panel. We indeed confirm that $j_{sd}=0$ at a special value $t'=t'_c$ (black line). In the left panel we show the gate evolution of the same parameters at $t'=t'_c$, with the 2CK conditions being satisfied here at $V_g=0$.


\begin{figure}[t!]
\includegraphics[width=1.0\linewidth]{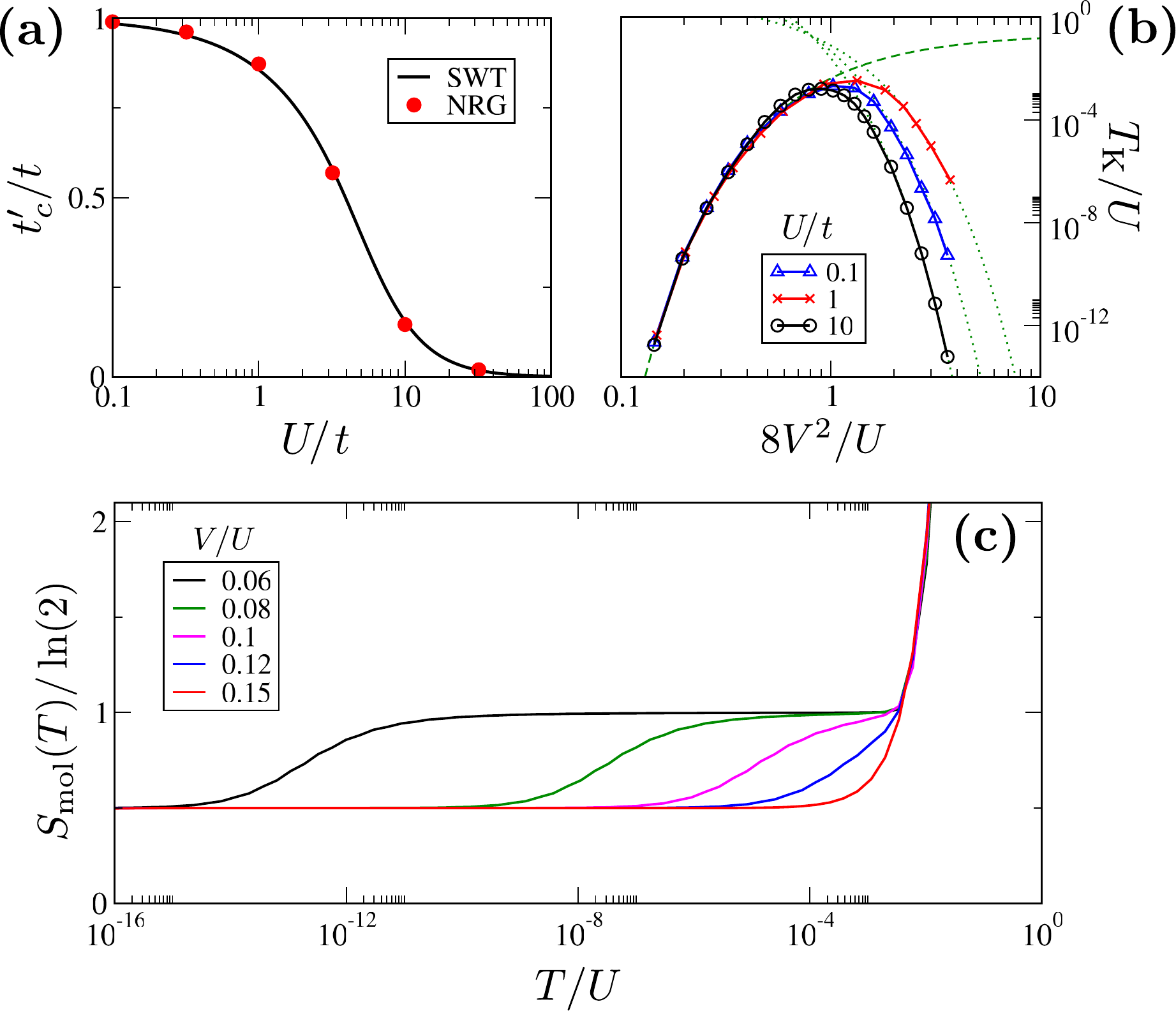}
  \caption{2CK critical point driven by QI. (a) Critical coupling $t'_c$ as a function of $U/t$, with NRG results (points) validating SWT predictions (line). (b) 2CK Kondo temperature $T_{\rm K}$ vs $8V^2/U\equiv J_K$ for different $U/t$ obtained by NRG. Dashed line is $T_{\rm K}/U\sim \exp[-4/J_K]$ valid for $J_K<1$ whereas the dotted lines show $T_{\rm K}/U\sim \exp[-a J_K]$ with $a \equiv a(U)\sim \mathcal{O}(1)$ for $J_K>1$.  (c) Entropy $S_{\rm mol}$ vs $T/U$ for different $V/U$ at the 2CK critical point for $U/t=10$, showing a residual $\tfrac{1}{2}\ln(2)$.}
  \label{fig:qcp}
\end{figure}

\textit{Non-perturbative solution: NRG.--} 
To confirm the existence of a 2CK critical point in this simple 5-site molecular cluster, we turn to the non-perturbative solution of the full molecular junction involving Eq.~\ref{eq:5site} using NRG \cite{wilson1975renormalization,*bulla2008numerical}, where we set $t=\tfrac{1}{2}$ and the conduction electron bandwidth $D=1$ from now on. Numerical results are presented in Fig.~\ref{fig:qcp}. In panel (a) we compare SWT predictions for the critical $t'_c$ with those obtained by NRG for different interaction strengths $U$, showing excellent agreement. In particular, we note that the 2CK critical point can be realized for \emph{any} finite $U$. Interestingly, we find that $t'_c\to t$ as $U\to 0$. The $U=0$ limit of Eq.~\ref{eq:5site} is studied in the SM \cite{SI}: we find $t'=t$ is a singular point of the non-interacting model with strictly decoupled molecular degrees of freedom that give a finite $T=0$ entropy and a QI-driven conductance node. With interactions switched on, the critical $t'_c$ is no longer at $t$ but we still find a residual $T=0$ entropy and a conductance node -- now characterizing the 2CK critical fixed point. Panel (c) shows the molecular contribution to the entropy $S_{\rm mol}$ as a function of $T$ at the critical point for different $V$. The critical point can be realized for any combination of $V$ and $U$ (in panel (c) we take fixed $U$), and in all cases we find $S_{\rm mol}=\tfrac{1}{2}\ln(2)$ for $T\ll T_{\rm K}$, with $T_{\rm K}$ the critical Kondo temperature. This unusual value for the entropy is a hallmark of the free Majorana fermion localized on the molecule at low temperatures at the 2CK critical point \cite{emery1992mapping,affleck1993exact}. For small molecule-lead coupling, $T_{\rm K}$ is small and we have an extended intermediate $\ln(2)$ plateau corresponding to the local moment regime of Eq.~\ref{eq:2ck}. Remarkably however, at larger $V$ the Kondo temperature can be boosted to large (non-universal) values and local moment physics is entirely eliminated. This scenario lies outside of the regime described by Eq.~\ref{eq:2ck}, 
suggesting that the interference giving rise to criticality is a topological feature of the geometry in 
Eq.~\ref{eq:5site}. In Fig.~\ref{fig:qcp}(b) we plot the evolution of the Kondo temperature with $8V^2/U \equiv J_K$ (where $J_K$ is the SWT Kondo coupling for a single Anderson impurity \cite{Hewson}), showing that a maximum value $T_{\rm K}\sim 10^{-2}U$ can be realized for all values of $U$ considered when $J_K \sim 1$. A weak-strong coupling duality \cite{kolf2007strong} is found on further increasing $J_K$ -- see dashed and dotted lines in Fig.~\ref{fig:qcp}(b). 
Note that the critical point is a non-Fermi liquid and as such is not perturbatively connected to the $U=0$ limit: even though the critical point can be realized at small $U$, we find that $T_{\rm K}\to 0$ as $U\to 0$.  


\begin{figure}[t!]
\includegraphics[width=0.89\linewidth]{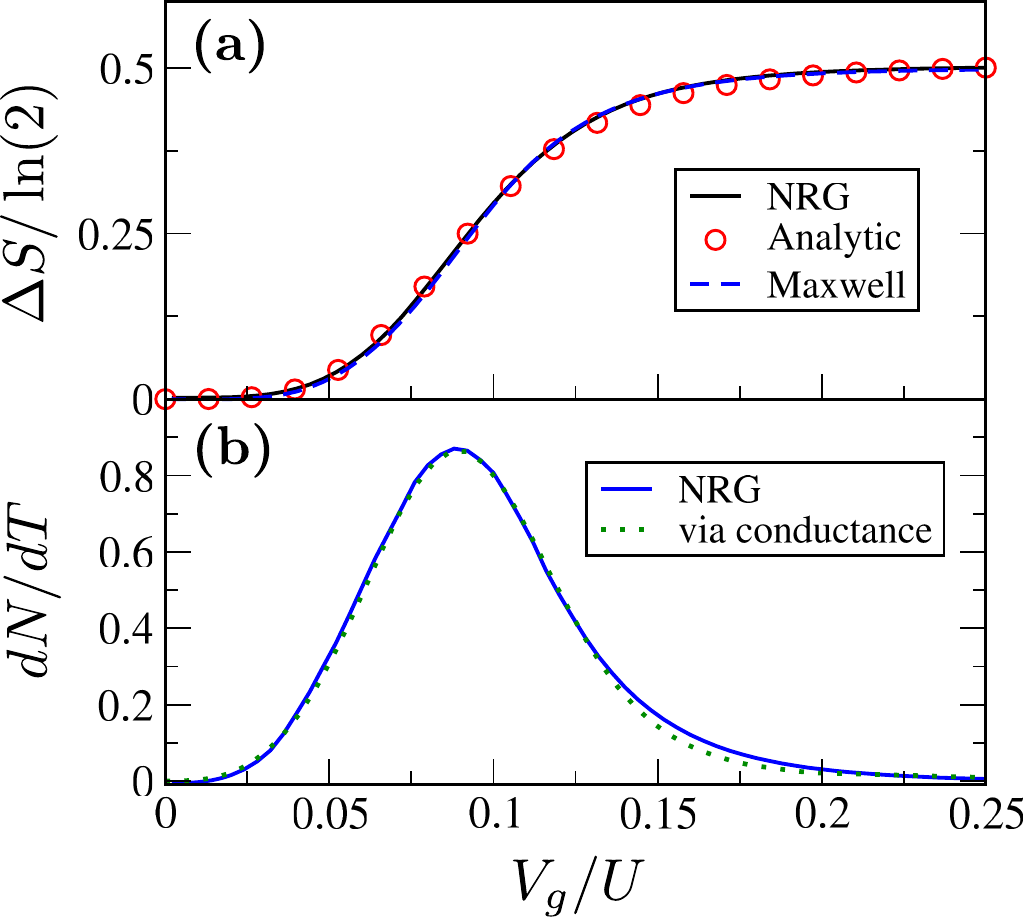}
  \caption{(a) Entropy change $\Delta S$ as the molecular junction is driven away from the critical point by increasing gate voltage $V_g$. NRG results (line) compared with analytic result Eq.~\ref{eq:entropy} (points). (b) $dN/dT$ from NRG (line), compared with prediction via conductance from Eq.~\ref{eq:dNdTcond} (dotted line). Dashed line in the top panel obtained by integrating $dN/dT$ over $V_g$. Plotted for $U/t=10$, $V/U=0.15$, $t'=t'_c$, $T=10^{-6} \ll T_{\rm K}$.
  }
  \label{fig:maxwell}
\end{figure}

\textit{Gate control and entropy measurement.--} 
With $t'$ tuned to the 2CK critical point at $t'_c$, we can vary the gate voltage $V_g$ in the vicinity of $V_g=0$. This perturbation drives the system away from the 2CK fixed point and towards a standard Kondo strong-coupling Fermi liquid (FL) state on the scale of $T^*$. From NRG we find \cite{SI},
\begin{eqnarray}\label{eq:tstar}
    T^* \sim V_g^4 \qquad :~~T^*\ll T_{\rm K}
\end{eqnarray} 
which holds in the universal critical regime. Along this FL crossover, physical properties are universal scaling functions of $T^*/T$ and hence $V_g/T^{1/4}$. For the pure 2CK model in this regime, bosonization methods give an exact result for the entropy change from the critical point \cite{emery1992mapping},
\begin{equation}\label{eq:entropy}
\hskip -0.2cm \Delta S\left(\frac{T^*}{T}\right) = \frac{T^*}{T}\left[\psi\left(\frac{1}{2} + \frac{T^*}{T}\right) - 1 \right] - \ln\left[\frac{1}{\sqrt{\pi}} \Gamma\left(\frac{1}{2} + \frac{T^*}{T}\right)\right]
\end{equation}
with $\Gamma$ ($\psi$) the gamma (digamma) function. The form of this crossover is entirely characteristic of the 2CK critical point \cite{Sela_2011,*Mitchell_2012a}. Using 
Eq.~\ref{eq:tstar}, this crossover can be achieved by fixing $T$ ($\ll T_K$) and detuning $V_g$ (which controls $T^*$). This is shown in the top panel of Fig.~\ref{fig:maxwell}, which compares NRG results for the junction (line) to Eq.~\ref{eq:entropy} (points).

Recent progress has been made in observing entropic signatures in nanoelectronics experiments, by exploiting local Maxwell relations which connect the entropy change for a process to measureable changes in the charge \cite{hartman2018direct,han2022fractional,child2022entropy}. Since the gate voltage $V_g$ couples to the total molecule charge $\hat{N}=\sum_{m}\hat{n}_m$, the change in entropy induced by scanning $V_g$ as in Fig.~\ref{fig:maxwell}(a) follows as $\Delta S = -\int dV_g ~dN/dT$. The quantity $dN/dT$ is shown in Fig.~\ref{fig:maxwell}(b). Application of the Maxwell relation yields the blue-dashed line in Fig.~\ref{fig:maxwell}(a), which agrees perfectly with the direct entropy calculation. We argue that the molecular system is well suited to this because $T_{\rm K}$ can be boosted to large values, meaning that the universal critical regime should be experimentally accessible. 


\begin{figure}[t!]
\includegraphics[width=1.05\linewidth]{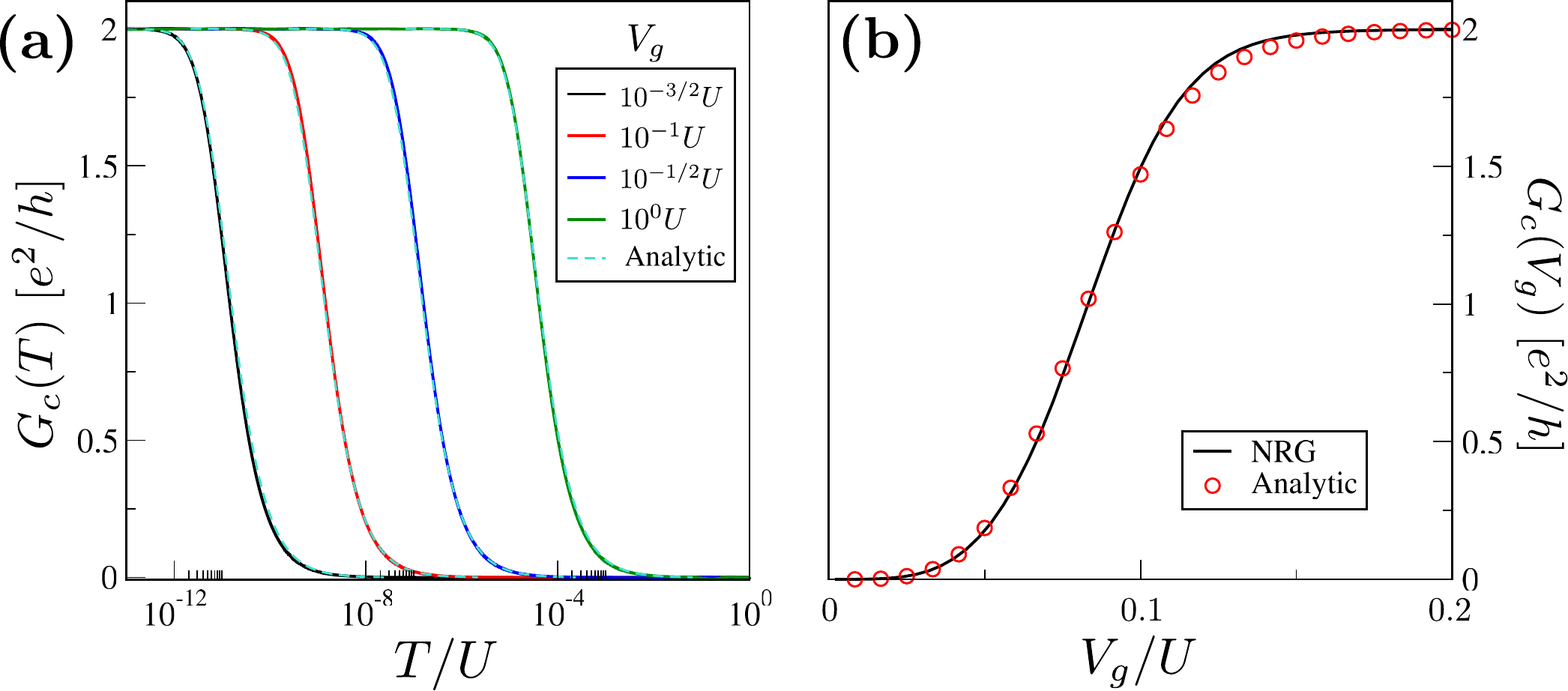}
  \caption{Series conductance along the FL crossover (a) as a function of temperature for different gate voltages; and (b) as a function of gate voltage at fixed $T=10^{-6}\ll T_{\rm K}$;  compared with Eq.~\ref{eq:cond}. Shown for $U/t=10$, $V/U=0.15$, $t'=t'_c$.
  }
  \label{fig:cond}
\end{figure}

\textit{Transport.--}
At the 2CK critical point, the series conductance through the molecular junction vanishes due to the many-body QI node. However, a nontrivial transport signature is picked up along the FL crossover by detuning the gate voltage. NRG results for the junction conductance $G_c(T)$ as a function of $T$ at fixed detuning $V_g$ are shown in Fig.~\ref{fig:cond}(a). The maximum conductance of $2e^2/h$ for a single electron transistor is recovered at low temperatures $T\ll T^*$ in all cases.   Fig.~\ref{fig:cond}(b) shows the gate evolution of the conductance $G_c(V_g)$ at fixed $T$ ($\ll T_{\rm K}$), and is the analogous plot to Fig.~\ref{fig:maxwell}(a). 

The exact solution of the pure 2CK model along the FL crossover \cite{emery1992mapping} yields a prediction for conductance \cite{Sela_2016},
\begin{equation}\label{eq:cond}
    G_c  \left ( \frac{T^*}{T}\right )=\frac{2e^2}{h}\times \left ( \frac{T^*}{T}\right )\psi'\left (\frac{1}{2}+ \frac{T^*}{T}\right ) \;,
\end{equation}
where $T^*$ depends on $V_g$ via Eq.~\ref{eq:tstar} and $\psi'$ is the trigamma function. This expression matches essentially perfectly with NRG data for the full molecular junction in Fig.~\ref{fig:cond}.

Finally, from the Maxwell relation $dN/dT=-dS/dV_g$ we can use Eqs.~\ref{eq:tstar}-\ref{eq:cond} to prove the exact conductance-charge relation \cite{han2022fractional} in the universal FL crossover regime,
\begin{equation}\label{eq:dNdTcond}
    \frac{dN}{dT} \sim \frac{V_g^3}{T}\left ( 1-\frac{Gc(V_g,T)}{2e^2/h}\right ) \;,
\end{equation}
meaning that experimental conductance data can be translated into $dN/dT$ (see Fig.~\ref{fig:maxwell}(b), dotted line) and then integrated to extract the entropy. 

\begin{figure}[t!]
\includegraphics[width=0.9\linewidth]{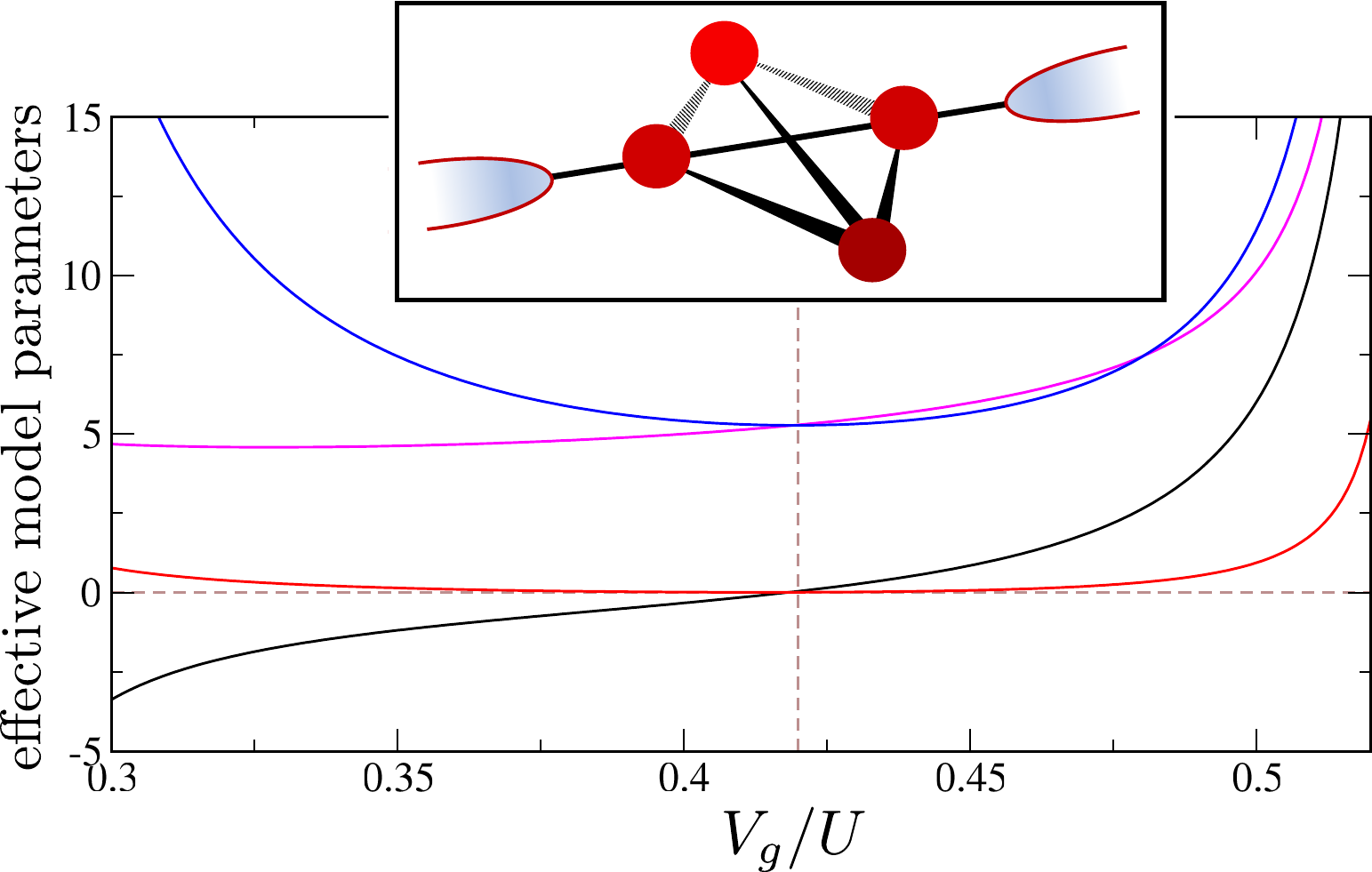}
  \caption{Inverse design of a 4-site molecular junction using GD optimization of effective SWT parameters to realize the QI-2CK condition at gate voltage $V_g/U=0.42$. Black: $j_{sd}$; red: $w_{sd}$; magenta: $j_{ss}$; blue: $j_{dd}$. Parameters given in \cite{SI}.
  }
  \label{fig:4site}
\end{figure}

\textit{Inverse design.--} The above results establish the existence of the QI-2CK effect in a simple molecular moiety with exact $ph$ and $sd$ symmetry. In a more general setting, however, we can use inverse design to search for candidate systems that satisfy the 2CK conditions. This can be done by setting up a loss function, for example $\mathcal{L}=j_{sd}^2+w_{sd}^2+(j_{ss}-j_{dd})^2$, which is minimum when the 2CK conditions on the effective model parameters are met. We then minimize this function with respect to the bare model parameters by gradient descent (GD). In practice this involves finding the derivatives of $j_{\alpha\beta}$ and $w_{\alpha\beta}$ with respect to $t_{mn}$ and $U_{mn}$, which can be achieved using differentiable programming techniques \cite{bartholomew2000automatic}. In the SM \cite{SI} we show that this can be implemented very efficiently within our improved SWT scheme.
Using this methodology, we could find a family of low-symmetry molecular junctions involving just 4 interacting sites \cite{SI}, a representative example of which is shown in Fig.~\ref{fig:4site}. By fine-tuning the gate voltage $V_g$ in this structure we predict 2CK criticality. We did not find any 2CK critical systems involving 1, 2, or 3 sites. 

A non-perturbative extension utilizing `differentiable NRG' \cite{rigo2022automatic} to optimize bare model parameters directly via GD could be used to bypass the SWT approximation.

\begin{acknowledgments}
\textit{Conclusion.--} 
The 2CK critical point can be realized by exploiting many-body QI effects in simple molecular junctions or coupled quantum dot devices, featuring a few tunnel-coupled, interacting orbitals. 
QI effects can be manipulated by tuning gate voltages to switch between a perfect node and perfect Kondo resonant transmission. 

Inverse design can be used to search automatically for systems displaying desired functionality. The molecular moieties we identified are not intended to be atomistic models of any real molecule. However, the inverse design approach could be integrated with chemical databases to search for realistic candidate molecular junctions \cite{raghunathan2022molecular}. In the SM we explore three such candidate molecules based on the simple design principles uncovered from our toy model studies \cite{SI}. Full \textit{ab initio} studies are left for future work. On the other hand, for artificial molecular junctions formed in semiconductor quantum dot devices, the simplest 4 or 5 site structures discussed here might be directly implementable.

Our results open the door to designer devices utilizing many-body QI effects. For example, simple structures exhibiting three-channel Kondo \cite{SI,iftikhar2018tunable} or two-impurity Kondo \cite{jones1988low,mitchell2012two,pouse2023quantum} effects, or lattice extensions describing non-Fermi liquid materials \cite{cox1996two}. Inverse design could be used to optimize performance of nanoscale transistors, rectifiers, spintronics devices and other quantum devices.

\vfill

 \textit{Acknowledgments.--} This work was supported by the Irish Research Council through the Laureate Award 2017/2018 grant IRCLA/2017/169 (AKM) and Science and Engineering Research Board, India
(SRG/2022/000495), (MTR/2022/000638), and IIT(ISM) Dhanbad [FRS(175)/2022-2023/PHYSICS] (SS). We thank Jonas Rigo for enlightening discussions.
\end{acknowledgments}


%

\end{document}


\title{\underline{Supplementary Material}\\\vskip 0.3cm Many-body quantum interference route to the two-channel Kondo effect:\\Inverse design for molecular junctions and quantum dot devices}


\author{Sudeshna Sen}
\affiliation{Department of Physics, IIT(ISM) Dhanbad, Dhanbad-826004, Jharkhand, India}
\author{Andrew K. Mitchell}
\affiliation{School of Physics, University College Dublin, Belfield, Dublin 4, Ireland}
\affiliation{Centre for Quantum Engineering, Science, and Technology, University College Dublin, Ireland}


\maketitle

\section{Microscopic models of\\molecular junctions}\label{sec:SI_models}
We describe molecular junctions (and coupled quantum dot devices constituting artificial molecular junctions) in terms of generalized quantum impurity models of the type,
\begin{equation}
    \hat{H}=\hat{H}_{\rm mol}+\hat{H}_{\rm leads}+\hat{H}_{\rm hyb} + \hat{H}_{\rm gate}
\end{equation}
where the part of the Hamiltonian $\hat{H}_{\rm mol}$ describing the isolated molecule is given by an extended Hubbard model,
\begin{equation}\label{eq:SI_Hmol}
    \hat{H}_{\rm mol} = \sum_{\sigma=\uparrow,\downarrow}\:\sum_{m,n} t_{mn}^{\phantom{\dagger}} d_{m\sigma}^{\dagger}d_{n\sigma}^{\phantom{\dagger}} + \tfrac{1}{2}\sum_{m,n} U_{mn} \hat{n}_m\hat{n}_n
\end{equation}
as given in Eq.~1 of the main text. Here $d_{m\sigma}^{(\dagger)}$ annihilates (creates) an electron on molecule orbital $m$ with spin $\sigma$ and $\hat{n}_m=\sum_{\sigma}\hat{n}_{m\sigma}$ is the total number operator for orbital $m$, where $\hat{n}_{m\sigma }= d_{m\sigma}^{\dagger}d_{m\sigma}^{\phantom{\dagger}}$.
The single-particle tunneling matrix elements between orbitals $m\ne n$ are given by $t_{mn}$ ($=t_{nm}^*$) whereas the onsite potentials are $\epsilon_m=t_{mm}+\tfrac{1}{2}U_{mm}$. The term $U_{mn}$ ($=U_{nm}$) embodies electronic interactions on the molecule. The standard local Hubbard interaction term $U_m \hat{n}_{m\uparrow}\hat{n}_{m\downarrow}$ between up and down spin electrons on a given site $m$ is already included in the diagonal part of the second term of Eq.~\ref{eq:SI_Hmol}, with $U_m\equiv U_{mm}$. The off-diagonal parts $U_{mn}$ for $m\ne n$ correspond to capacitive (Coulomb) repulsion between different sites. In principle one can also add spin-orbit coupling terms, anisotropies, and exchange interactions. The formalism presented below can be straightforwardly adapted to treat any $\hat{H}_{\rm mol}$.

The source ($\alpha=s$) and drain ($\alpha=d$) leads are taken to be thermal reservoirs of non-interacting conduction electrons, $\hat{H}_{\rm leads}=\sum_{\alpha} \hat{H}_{\rm leads}^{\alpha}$ with 
\begin{equation}
    \hat{H}_{\rm leads}^{\alpha}=\sum_{k,\sigma}\epsilon_{k}^{\phantom{\dagger}}c_{\alpha k \sigma}^{\dagger} c_{\alpha k \sigma}^{\phantom{\dagger}}
\end{equation}
where for simplicity we take the dispersion $\epsilon_k$ to be independent of spin $\sigma$ and equivalent for both leads $\alpha$ (although again this can be straightforwardly generalized without affecting the following). 

The molecule frontier orbital $d_{r_{\alpha} \sigma}$ couples to the conduction electrons of lead $\alpha$ according to the hybridization term
\begin{equation}
    \hat{H}_{\rm hyb}=\sum_{\alpha,k,\sigma}V_{\alpha k}^{\phantom{\dagger}}(d_{r_{\alpha}\sigma}^{\dagger}c_{\alpha k \sigma}^{\phantom{\dagger}}+{\rm H.c.}) \;.
\end{equation} 
We now define the local lead orbitals at the molecule position as $c_{\alpha\sigma}=\tfrac{1}{V_{\alpha}}\sum_{k}V_{\alpha k} c_{\alpha \sigma k}$ with $V_{\alpha}^2=\sum_k V_{\alpha k}^2$ such that $\hat{H}_{\rm hyb}=\sum_{\alpha,\sigma}V_{\alpha}(d_{r_{\alpha}\sigma}^{\dagger}c_{\alpha\sigma}^{\phantom{\dagger}}+{\rm H.c.})$. 

The free lead density of states at the junction position (taken to the same for both leads and independent of spin) is $\rho(\omega)=-\tfrac{1}{\pi}{\rm Im}~G_{\rm lead}^0(\omega)$ where $G_{\rm lead}^0(\omega)=\langle\langle c_{\alpha\sigma}^{\phantom{\dagger}} ; c_{\alpha\sigma}^{\dagger} \rangle\rangle$ is the free Green's function for the local lead orbital $c_{\alpha\sigma}$ to which the molecule couples. For simplicity we use $\rho(\omega)=\rho_0\theta(D-|\omega|)$ such that the lead density of states is constant $\rho_0=1/2D$ in a band of half-width $D$. Throughout this work we set $D=1$.

The gate voltage $V_g$ allows to tune the number of electrons on the molecule via $\hat{H}_{\rm gate}=V_g\sum_m \hat{n}_m$. 


\subsection{Interacting nanowire (Hubbard chains)}\label{sec:si_1dhub_model}
In the main text we discuss a particular limiting case of Eq.~\ref{eq:SI_Hmol} corresponding to an $M$-site 1d Hubbard chain,
\begin{equation}\label{eq:SI_1dhub}
   \hskip -0.2cm \hat{H}_{\rm mol}=t\sum_{m=1}^{M-1}\sum_{\sigma} \left ( d_{m\sigma}^{\dagger}d_{m+1 \sigma}^{\phantom{\dagger}} + {\rm H.c.} \right )  + \sum_{m=1}^{M}\left (\epsilon\hat{n}_m + U \hat{n}_{m\uparrow}\hat{n}_{m\downarrow} \right )
\end{equation}
which describes an interacting nanowire (for example the $\pi$-system of a conjugated hydrocarbon polymer \cite{lafferentz2009conductance}). We take constant nearest-neighbour tunneling $t$, constant onsite potentials $\epsilon$ and constant local Hubbard repulsion $U$. The source and drain leads are connected to the ends of the nanowire at sites $r_{s}=1$ and $r_{d}=M$, such that
\begin{equation}\label{eq:SI_1dhubhyb}
H_{\rm hyb} = \sum_{\sigma} \left ( V_s^{\phantom{\dagger}} c_{s\sigma}^{\dagger}d_{1\sigma}^{\phantom{\dagger}} + V_d^{\phantom{\dagger}} c_{d\sigma}^{\dagger}d_{M\sigma}^{\phantom{\dagger}} + {\rm H.c.} \right )\;.
\end{equation}
The low-energy effective model for this system with odd chain length $M$ is discussed below in Sec.~\ref{sec:swtoddchain}.


\subsection{Symmetric 5-site cluster}\label{sec:si_5site_model}
Introducing additional tunnel-couplings between different sites of the above 1d Hubbard chain clearly produces greater complexity and, in particular, more possibilities for QI effects.

However, if we want exact $ph$ symmetry then the molecule connectivity must be compatible with a bipartite graph -- which in the present context simply means that there should not be any odd-membered loops. The simplest system with an
odd total number of sites with $sd$-symmetry to contain
multiple paths through the molecule, while also realizing exact particle hole symmetry at $V_g = 0$, is the
$M = 5$ chain with next-next-nearest-neighbour tunneling connecting sites $1\leftrightarrow 4$ and $2\leftrightarrow 5$,
\begin{equation}
\begin{split}\label{eq:SI_5site}
    H_{\rm mol} =& \sum_{m=1}^5 \left ( -\tfrac{1}{2}U \hat{n}_m + U \hat{n}_{m\uparrow} \hat{n}_{m\downarrow} \right ) \\ 
    &+ t\sum_{m=1}^4 \sum_{\sigma} \left( d_{m\sigma}^{\dagger} d_{m+1\sigma}^{\phantom{\dagger}}+{\rm H.c.} \right ) \\
    &+ t'\sum_{\sigma}\left( d_{1\sigma}^{\dagger} d_{4\sigma}^{\phantom{\dagger}}+ d_{2\sigma}^{\dagger} d_{5\sigma}^{\phantom{\dagger}} +{\rm H.c.} \right )
\end{split}
\end{equation}
which is equivalent to Eq.~3 of the main text. We attach leads to sites $1$ and $5$ via Eq.~\ref{eq:SI_1dhubhyb}.
We discuss the low-energy effective model for this system below in Sec.~\ref{sec:si_5site}


\subsection{Low-symmetry 4-site cluster}\label{sec:si_4site_model}
Finally we consider a general low-symmetry structure with just $M=4$ interacting sites. We simplify to local Hubbard interactions:
\begin{equation}\label{eq:si_4site}
    \hat{H}_{\rm mol} = \sum_{m,n}\sum_{\sigma} t_{mn}^{\phantom{\dagger}} d_{m\sigma}^{\dagger}d_{n\sigma}^{\phantom{\dagger}} + \sum_{m} U_{m} \hat{n}_{m\uparrow}\hat{n}_{m\downarrow}
\end{equation}
Leads are attached to sites $1$ and $4$ via Eq.~\ref{eq:SI_1dhubhyb}.
We discuss gradient descent optimization of this system to satisfy the QI-2CK conditions in Sec.~\ref{sec:si_inv}.

The parameters used in Fig.~5 of the main text are $t_{11} \simeq 0.697$,~~$t_{22} \simeq 0.642$,~~$t_{33} \simeq 0.480$,~~$t_{44} \simeq 0.253$,~~$V_g=0.42$,~~$t_{12} \simeq 0.102$,~~$t_{13} \simeq 0.053$,~~$t_{14} \simeq 0.1$,~~$t_{23}\simeq 0.109$,~~$t_{24}\simeq 0.046$,~~$t_{34}\simeq 0.169$.


\section{Generalized numerical Schrieffer-Wolff transformation}\label{sec:SI_SWT}
The Schrieffer-Wolff transformation (SWT) was first applied to the mapping between the Anderson impurity model (AIM) and the Kondo model \cite{schrieffer1966relation}. For the AIM it involves projecting onto the impurity spin states by eliminating virtual excitations and folding in their contribution to effective model parameters. The SWT is in essence a single-step renormalisation group (RG) procedure. Typically the mapping is done perturbatively to second order in the impurity-bath hybridization, which is then equivalent to performing second-order Brillouin-Wigner perturbation theory (BWPT) \cite{Hewson}. For the AIM-to-Kondo mapping this can be done straightforwardly and analytic expressions for the Kondo model parameters are easily obtained.

However, in the present context of complex molecular junctions, an exact analytic treatment of the SWT is in general not possible, and we must resort to a \emph{numerical} evaluation of the effective model parameters \cite{pedersen2014quantum,mitchell2017kondo}. Such a generalized numerical SWT is an ideal tool for efficiently describing many-body quantum interference (QI) effects. It can also be integrated with machine learning (ML) algorithms for inverse design, by identifying candidate molecular junctions that have specific desired effective model properties, as shown below.

Before proceeding we make a comment on the validity of the SWT. The RG character of quantum impurity systems means that a second-order SWT typically yields the correct operator structure of the low-energy effective model, even if the precise value of the derived parameters is not exact \cite{rigo2020machine}. However, even infinite-order SWT \cite{chan2004exact} neglects retardation effects from the conduction electrons resulting in bandwidth renormalization \cite{haldane1978theory}, and more sophisticated non-perturbative techniques such as the numerical renormalization group \cite{bulla2008numerical} (NRG) are ultimately required. On the other hand, in this work we demonstrate that QI effects are very well captured by SWT (even quantitatively) when compared with NRG.


\subsection{Formalism}
In this work we are interested in deriving generalized two-channel Kondo (2CK) models of the type Eq.~2 from the main paper. Such a description arises in two-lead molecular junctions as the low-energy effective model when the isolated molecule ground state manifold comprises a unique spin-doublet \cite{pedersen2014quantum,mitchell2017kondo}. The structure of the effective model is constrained by $U(1)$ charge conservation and $SU(2)$ spin symmetry. The spin-doublet ground state of the molecule implies that the ground state hosts an odd number $N$ of electrons, and that tunneling to and from the leads involves $N\pm 1$ even-electron sectors with integer spin.

With $\hat{H}'_{\rm mol}=\hat{H}_{\rm mol}+\hat{H}_{\rm gate}$ taken to include gate voltage effects on the isolated molecule, we define a set of exact eigenstates $\{ |\psi^{n;S^z}_j \rangle \}$ satisfying the Schr\"odinger equation $\hat{H}'_{\rm mol} |\psi^{n;S^z}_j \rangle = E_{j}^{n;S^z}|\psi^{n;S^z}_j \rangle$, where $n$ is the number of electrons on the isolated molecule, $S^z$ is the total molecule spin projection, and $j$ is an index distinguishing molecule states with the same $n$ and $S^z$. The ground spin doublet states are specified by $|\psi^{N;\sigma}_0 \rangle$ with $\sigma=\uparrow,\downarrow$, and the corresponding ground state energy is $E_{\rm gs}\equiv E_{0}^{N;\uparrow}=E_{0}^{N;\downarrow}$.

Eigenstates of $\hat{H}'_{\rm mol}$ can be expressed in the (computational) many-body product-state basis $\{ |\phi^{n;S^z}_{k}\rangle\}$, which are labelled by the same quantum numbers, viz:
\begin{equation}\label{eq:SI_eig}
    |\psi^{n;S^z}_j\rangle = \sum_k U^{n;S^z}_{k,j}  |\phi^{n;S^z}_k\rangle \;.
\end{equation}
For notational convenience we define product basis vectors $\vec{\phi}^{n,S^z}  = \left ( |\phi^{n;S^z}_0\rangle, ~|\phi^{n;S^z}_1\rangle, ~|\phi^{n;S^z}_2\rangle,~ ... \right )$ and eigenbasis vectors 
$\vec{\psi}^{n,S^z}  = \left ( |\psi^{n;S^z}_0\rangle, ~|\psi^{n;S^z}_1\rangle, ~|\psi^{n;S^z}_2\rangle,~ ... \right )$. These are related by Eq.~\ref{eq:SI_eig} as $\vec{\psi}^{n,S^z} = \vec{\phi}^{n,S^z} \boldsymbol{U}^{n;S^z}$ in terms of the transformation matrices $\boldsymbol{U}^{n;S^z}$ with elements $U^{n;S^z}_{k,j}$. The Hamiltonian matrix for the isolated but interacting molecule in the computational product basis of a given quantum number subspace can then be written as 
$\boldsymbol{H}_{\phi}^{n;S^z} = \left (\vec{\phi}^{n,S^z} \right )^{\dagger} \hat{H}'_{\rm mol} ~\vec{\phi}^{n,S^z}$, whereas in the 
eigenbasis we have 
$\boldsymbol{H}_{\psi}^{n;S^z} = \left (\vec{\psi}^{n,S^z} \right )^{\dagger} \hat{H}'_{\rm mol} ~\vec{\psi}^{n,S^z} = \left (\boldsymbol{U}^{n;S^z} \right )^{\dagger} \boldsymbol{H}_{\phi}^{n;S^z} \boldsymbol{U}^{n;S^z} \equiv {\rm diag}\left (E_0^{n,S^z},E_1^{n,S^z},E_2^{n,S^z},... \right )$. These expressions hold separately in each subspace. 

Finally, we will find it useful to define the vector of expansion coefficients for the $N$-electron \emph{ground state} with spin $\sigma$ as $\vec{U}^{N;\sigma}_{\rm gs}= \left (U^{N;\sigma}_{0,0},~U^{N;\sigma}_{1,0},~U^{N;\sigma}_{2,0},~...\right )^T$ such that  $|\psi^{N;\sigma}_0\rangle=  \vec{\phi}^{N;\sigma} \cdot \vec{U}^{N;\sigma}_{\rm gs}$. 
We may also denote $\vec{\phi}_{\rm gs} \equiv  \vec{\phi}^{N;\uparrow}$ and $|\psi_{\rm gs}\rangle \equiv  |\psi^{N;\uparrow}_0\rangle$ and $\vec{U}_{\rm gs} \equiv \vec{U}^{N;\uparrow}_{\rm gs}$ for notational compactness where applicable.\\


With the above preliminaries established, we now use BWPT to project the full Hamiltonian $\hat{H}=\hat{H}_{\rm leads}+\hat{H}_{\rm mol}'+\hat{H}_{\rm hyb}$ onto the low-energy manifold containing the spin-doublet molecule ground states, by perturbatively eliminating excitations to higher-lying molecule states. We define the projector $\hat{P}=\sum_{\sigma=\uparrow,\downarrow} |\psi_0^{N,\sigma}\rangle\langle \psi_0^{N,\sigma}|$ for the ground states to be kept and $\hat{Q}=\hat{1}-\hat{P}$ for the excited states to be discarded in our effective model. To second order in $\hat{H}_{\rm hyb}$ we obtain an effective Hamiltonian,
\begin{equation}\label{eq:SI_bwpt}
    \hat{H}_{\rm eff} = \hat{P}\hat{H}\hat{P} + \hat{P} \hat{H}_{\rm hyb} \hat{Q} (E_{\rm gs} - \hat{H}'_{\rm mol})^{-1} \hat{Q} \hat{H}_{\rm hyb} \hat{P} \;.
\end{equation}
With the specific form of the hybridization Hamiltonian,
\begin{equation}\label{eq:SI_hyb}
    \hat{H}_{\rm hyb} = \sum_{\substack{s=\uparrow,\downarrow \\ \alpha=s,d}}V_{\alpha}^{\phantom{\dagger}}  \left ( c_{\alpha s}^{\dagger} d_{r_{\alpha} s}^{\phantom{\dagger}} + d_{r_{\alpha} s}^{\dagger}c_{\alpha s}^{\phantom{\dagger}} \right ) \;,
\end{equation}
the effective model takes the form,
\begin{equation}\label{eq:SI_HeffA}
    \hat{H}_{\rm eff} = \hat{H}_{\rm leads} + \sum \left (V_{\alpha}V_{\beta} ~A{\substack{\sigma\sigma' \\ s,s' \\ \alpha,\beta}} \right ) c_{\beta s'}^{\dagger}c_{\alpha s}^{\phantom{\dagger}}
 |\psi_0^{N;\sigma'}\rangle \langle \psi_0^{N;\sigma}| \;,
\end{equation}
where a sum over all indices of the tensor $A$ is implied. Charge conservation, $SU(2)$ spin symmetry and Hermiticity impose constraints on the the tensor $A$. Together with the definitions,
\begin{align}
    \hat{\mathbb{S}}^+ &= |\psi_0^{N;\uparrow}\rangle \langle \psi_0^{N;\downarrow}| \;\\
   \hat{\mathbb{S}}^- &= |\psi_0^{N;\downarrow}\rangle \langle \psi_0^{N;\uparrow}| \;\\
    \hat{\mathbb{S}}^{z\;} &= \tfrac{1}{2} \left [|\psi_0^{N;\uparrow}\rangle \langle \psi_0^{N;\uparrow}| - |\psi_0^{N;\downarrow}\rangle \langle \psi_0^{N;\downarrow}|  \right ]\;
\end{align}
for the retained molecule states and 
\begin{equation}
    \hat{\vec{s}}_{\alpha\beta} = \tfrac{1}{2} \sum_{ss'} c_{\beta s'}^{\dagger} \vec{\boldsymbol{\sigma}}_{s's} ~c_{\alpha s}^{\phantom{\dagger}} \;,
\end{equation}
for the conduction electrons, we obtain the effective generalized 2CK model \cite{pedersen2014quantum,mitchell2017kondo}, Eq.~2 of the main paper:
\begin{equation}\label{eq:SI_2ck}
    \hat{H}_{\rm eff} = \hat{H}_{\rm leads} + \sum_{\alpha\beta} \left [J_{\alpha\beta\;}\hat{\vec{\mathbb{S}}}\cdot \hat{\vec{s}}_{\alpha\beta} + W_{\alpha\beta\;}^{\phantom{\dagger}}\sum_{\sigma} c_{\beta \sigma}^{\dagger} c_{\alpha \sigma}^{\phantom{\dagger}} \right ] \;.
\end{equation}
In the following we consider the rescaled couplings $j_{\alpha\beta}$ and $w_{\alpha\beta}$ defined by $J_{\alpha\beta}=V_{\alpha}V_{\beta} j_{\alpha\beta}$ and $W_{\alpha\beta}=V_{\alpha}V_{\beta} w_{\alpha\beta}$, which follow from the factorized form of the couplings up to second order in Eq.~\ref{eq:SI_HeffA}. It immediately follows that:
\begin{align}
    j_{\alpha\beta} &= 2\left ( A{\substack{\uparrow\uparrow \\ \uparrow\uparrow \\ \alpha,\beta}} - A{\substack{\uparrow\uparrow \\ \downarrow\downarrow \\ \alpha,\beta}}\right ) \; \label{eq:SI_j}\\ 
    w_{\alpha\beta} &= \tfrac{1}{2}\left ( A{\substack{\uparrow\uparrow \\ \uparrow\uparrow \\ \alpha,\beta}} + A{\substack{\uparrow\uparrow \\ \downarrow\downarrow \\ \alpha,\beta}}\right )  \; \label{eq:SI_w}
\end{align}
in terms of the amplitudes appearing in Eq.~\ref{eq:SI_HeffA}. There are other equivalent forms which follow from the symmetry transformations but we shall use these ones. \\

The final step is of course to calculate the amplitudes $A$. This is done by inserting Eq.~\ref{eq:SI_hyb} into Eq.~\ref{eq:SI_bwpt} and comparing with Eq.~\ref{eq:SI_HeffA}. The amplitudes decompose into particle and hole tunneling processes
\begin{equation}\label{eq:SI_A}
    A{\substack{\sigma\sigma' \\ s,s' \\ \alpha,\beta}}  = p{\substack{\sigma\sigma' \\ s,s' \\ \alpha,\beta}}  - h{\substack{\sigma\sigma' \\ s,s' \\ \alpha,\beta}} 
\end{equation}
where
\begin{equation}
   p{\substack{\sigma\sigma' \\ s,s' \\ \alpha,\beta}} = 
   \sum_j \frac{ \langle \psi^{N;\sigma'}_0 | d_{r_{\beta} s'}^{\phantom{\dagger}} | \psi^{N+1;\sigma+s}_j \rangle \langle \psi^{N+1;\sigma+s}_j | d_{r_{\alpha} s}^{\dagger} | \psi^{N;\sigma}_0 \rangle }{ E_{\rm gs} - E^{N+1;\sigma+s}_j} 
\end{equation}
for the particle process and 
\begin{equation}
   h{\substack{\sigma\sigma' \\ s,s' \\ \alpha,\beta}} = 
   \sum_j \frac{ \langle \psi^{N;\sigma'}_0 | d_{r_{\beta} s'}^{\dagger} | \psi^{N-1;\sigma-s}_j \rangle \langle \psi^{N-1;\sigma-s}_j | d_{r_{\alpha} s}^{\phantom{\dagger}}| \psi^{N;\sigma}_0 \rangle }{ E_{\rm gs} - E^{N-1;\sigma-s}_j} 
\end{equation}
for the hole process. Note that spin conservation requires $\sigma+s=\sigma'+s'$ for particle tunneling and $\sigma-s=\sigma'-s'$ for hole tunneling. \\

Numerical calculations can be made dramatically more efficient by exploiting the following observation: we may express particle and hole amplitudes in matrix form, viz:
\begin{align}
    p{\substack{\sigma\sigma' \\ s,s' \\ \alpha,\beta}} &= \left [ \left(\boldsymbol{X}^{N;\sigma'}_{\beta s'+}\right)^{\dagger} \left [ E_{\rm gs} \boldsymbol{I} - \boldsymbol{H}_{\psi}^{N+1;\sigma+s} \right ]^{-1} \left (\boldsymbol{X}^{N;\sigma}_{\alpha s +}\right ) \right ]_{00}\label{eq:SI_p} \\
    h{\substack{\sigma\sigma' \\ s,s' \\ \alpha,\beta}} &= \left [ \left(\boldsymbol{X}^{N;\sigma'}_{\beta s'-}\right)^{\dagger} \left [ E_{\rm gs} \boldsymbol{I} - \boldsymbol{H}_{\psi}^{N-1;\sigma-s} \right ]^{-1} \left (\boldsymbol{X}^{N;\sigma}_{\alpha s-}\right ) \right ]_{00} \label{eq:SI_h} 
\end{align}
where $\boldsymbol{X}^{N;\sigma}_{\alpha s \pm}= \left (\vec{\psi}^{N\pm 1,\sigma \pm s} \right )^{\dagger} d_{r_{\alpha}s}^{(\dagger)} ~\vec{\psi}^{N;\sigma}$ 
is evaluated in the eigenbasis. Switching now to the product basis, we write $\boldsymbol{X}_{\alpha s \pm}^{N;\sigma} =  \left (\boldsymbol{U}^{N \pm 1;\sigma \pm s} \right )^{\dagger} \boldsymbol{\mathbb{X}}^{N;\sigma}_{\alpha s \pm} \;\boldsymbol{U}^{N;\sigma}$, where $\boldsymbol{\mathbb{X}}^{N;\sigma}_{\alpha s \pm}= \left (\vec{\phi}^{N\pm 1,\sigma \pm s} \right )^{\dagger} d_{r_{\alpha}s}^{(\dagger)} ~\vec{\phi}^{N;\sigma}$.\\
Inserting into Eqs.~\ref{eq:SI_p} and \ref{eq:SI_h} we finally obtain,
\begin{align}
p{\substack{\sigma\sigma' \\ s,s' \\ \alpha,\beta}} &= \left (\vec{U}^{N;\sigma'}_{\rm gs} \right )^{\dagger} \boldsymbol{\mathbb{M}}{\substack{\sigma\sigma' \\ s,s' \\ \alpha,\beta}}(+)~ \vec{U}^{N;\sigma}_{\rm gs} \label{eq:SI_pM}\\
h{\substack{\sigma\sigma' \\ s,s' \\ \alpha,\beta}} &= \left (\vec{U}^{N;\sigma'}_{\rm gs} \right )^{\dagger} \boldsymbol{\mathbb{M}}{\substack{\sigma\sigma' \\ s,s' \\ \alpha,\beta}}(-)~ \vec{U}^{N;\sigma}_{\rm gs} \label{eq:SI_hM}
\end{align}
with 
\begin{align}\label{eq:SI_M}
\boldsymbol{\mathbb{M}}{\substack{\sigma\sigma' \\ s,s' \\ \alpha,\beta}}(\pm)=\left(\boldsymbol{\mathbb{X}}^{N;\sigma'}_{\beta s'\pm}\right)^{\dagger} \left [ E_{\rm gs} \boldsymbol{I} - \boldsymbol{H}_{\phi}^{N\pm 1;\sigma\pm s} \right ]^{-1} \left (\boldsymbol{\mathbb{X}}^{N;\sigma}_{\alpha s \pm}\right )
\end{align}
and where $\vec{U}^{N;\sigma}_{\rm gs}$ is the vector of expansion coefficients for the \emph{ground state} wavefunction only, as above.\\

The beauty of Eqs.~\ref{eq:SI_pM}, \ref{eq:SI_hM}, \ref{eq:SI_M} is that only the \textit{ground} eigenstate $|\psi^{N;\sigma}_0\rangle$ and the ground state energy $E_{\rm gs}$ of the isolated molecule are required: one does \textit{not} need to fully diagonalize the entire molecule. The matrix $\mathbb{M}$ is simply evaluated in the original product basis. This is highly efficient since one can use eigensolvers targeting just the ground state for the calculation, meaning that large systems can be dealt with cheaply. This in turn permits high throughput calculations which can be integrated within ML routines for inverse design.\\

Although the above formulation is entirely general, we note from Eqs.~\ref{eq:SI_j}, \ref{eq:SI_w} that in the present context, the 2CK mapping requires only a few elements of the full tensor $A$. Specifically, we may set $\sigma=\sigma'=\uparrow$, which we assume from now on. Furthermore, spin conservation then implies that $s=s'$. Therefore we abbreviate our full $A$, $p$ and $h$ tensors to,
\begin{equation}
A{\substack{\uparrow\uparrow \\ s,s \\ \alpha,\beta}}  \to  A^{s\alpha\beta} = p^{s\alpha\beta} - h^{s\alpha\beta} \;.
\end{equation}
Likewise we abbreviate,
\begin{equation}
\boldsymbol{\mathbb{M}}{\substack{\sigma\sigma' \\ s,s' \\ \alpha,\beta}}(\pm) \to \boldsymbol{\mathbb{M}}^{s\alpha\beta}_{\pm} \;,
\end{equation}
as used in the main paper. 
Thus we may write,
\begin{align}
p^{s\alpha\beta} &= \left (\vec{U}^{N;\uparrow}_{\rm gs} \right )^{\dagger} \boldsymbol{\mathbb{M}}^{s\alpha\beta}_+~ \left ( \vec{U}^{N;\uparrow}_{\rm gs} \right )\\
h^{s\alpha\beta} &= \left (\vec{U}^{N;\uparrow}_{\rm gs} \right )^{\dagger} \boldsymbol{\mathbb{M}}^{s\alpha\beta}_-~ \left (\vec{U}^{N;\uparrow}_{\rm gs} \right )
\end{align}
\begin{align}
\boldsymbol{\mathbb{M}}^{s\alpha\beta}_\pm=\left(\boldsymbol{\mathbb{X}}^{N;\uparrow}_{\beta s\pm}\right)^{\dagger} \left [ E_{\rm gs} \boldsymbol{I} - \boldsymbol{H}_{\phi}^{N\pm 1;\tfrac{1}{2}\pm s} \right ]^{-1} \left (\boldsymbol{\mathbb{X}}^{N;\uparrow}_{\alpha s\pm}\right )
\end{align}
\begin{align}
\boldsymbol{\mathbb{X}}^{N;\uparrow}_{\alpha s \pm}= \left (\vec{\phi}^{N\pm 1,\tfrac{1}{2} \pm s} \right )^{\dagger} d_{r_{\alpha} s}^{(\dagger)} ~\left ( \vec{\phi}^{N;\uparrow}\right )
\end{align}


\subsection{SWT results for odd Hubbard chains}\label{sec:swtoddchain}
As a simple application, we consider the SWT and resulting effective 2CK model parameters as a function of gate voltage $V_g$ for chains of $M$ interacting sites with on-site Coulomb repulsion $U$, connected by nearest-neighbour hoppings $t$. The Hamiltonian is given by Eq.~\ref{eq:SI_1dhub}, and we take $\epsilon=-U/2$ here so that the system possesses exact particle-hole symmetry when $V_g=0$. Leads are connected to either end of the chain, Eq.~\ref{eq:SI_1dhubhyb}. For odd $M$, the molecule supports a unique spin-doublet ground state for gate voltages around $V_g=0$ (the Coulomb blockade transitions to $M\pm 1$ electron states depend on $U$, and $t$). When the molecule hosts a spin-doublet ground state, the effective 2CK model, Eq.~\ref{eq:SI_2ck}, describes the low-energy behavior \cite{mitchell2011two}. 

\begin{figure}[t]
\includegraphics[width=\linewidth]{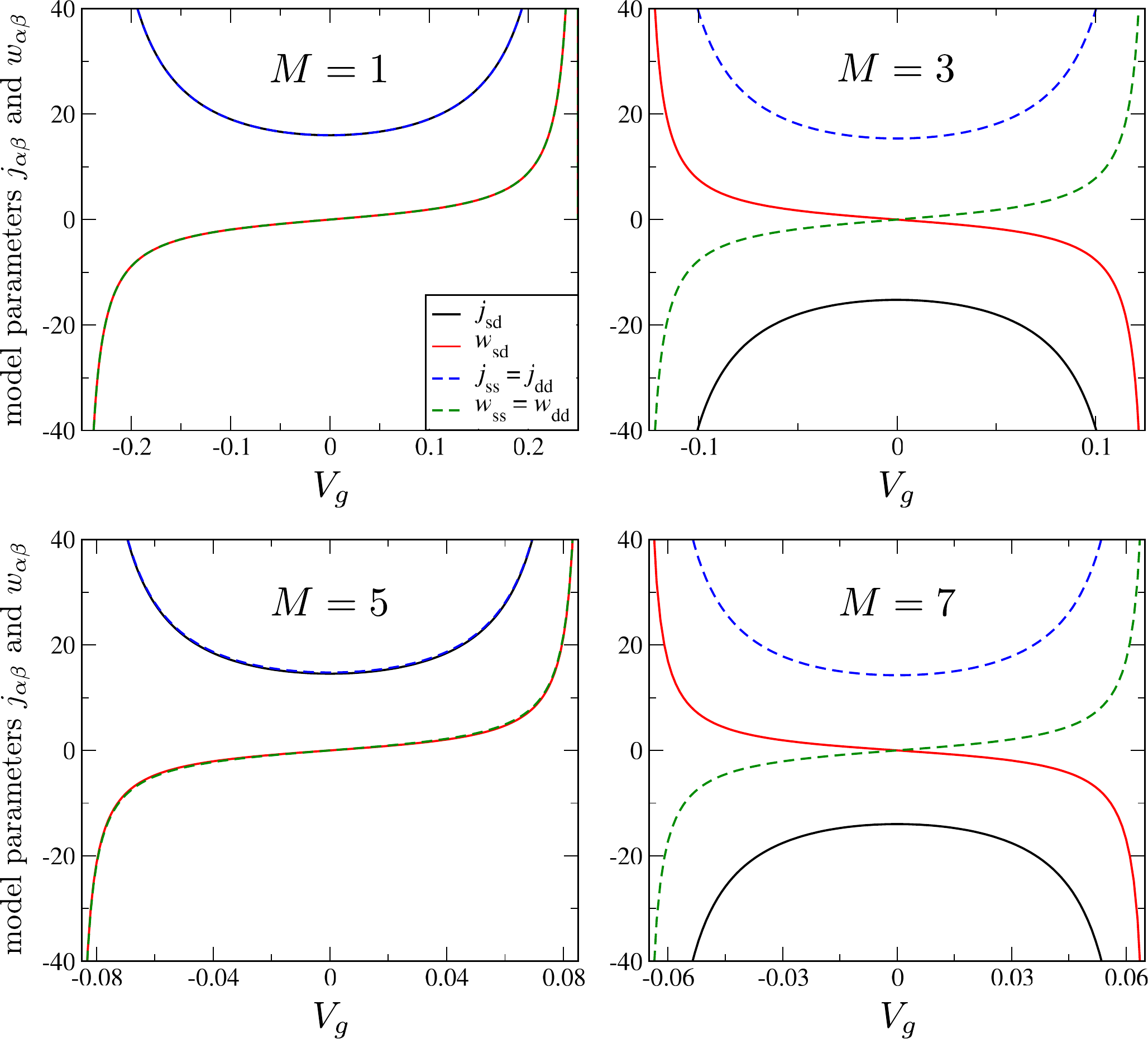}
  \caption{Effective model parameters $j_{\alpha\beta}$ and $w_{\alpha\beta}$ as a function of gate voltage $V_g$ from SWT calculations on chains of length $M=1,3,5,7$ sites, with local interaction $U$ and nearest-neighbour hopping $t$. In this example we have taken $U/t=1$ and set $t=0.5$. Source and drain leads are connected to sites $1$ and $M$. Kondo exchange interactions $j_{ss}$ and $j_{dd}$ (blue dashed lines) are equal and antiferromagnetic throughout the entire gate voltage range shown where the many-body molecule ground state is a spin-doublet. Potential scatterings $w_{ss}$, $w_{dd}$, and $w_{sd}$ vanish at the particle-hole symmetric point $V_g=0$ in all cases. The sign of $j_{sd}$ is seen to alternate, being antiferromagnetic for $M=4m+1$ and ferromagnetic for $M=4m+3$.
  }
  \label{fig:SI_chains}
\end{figure}

Fig.~\ref{fig:SI_chains} shows the effective 2CK model parameters $j_{\alpha\beta}$ and $w_{\alpha\beta}$ as a function of gate voltage $V_g$ for odd $M=1,3,5,7$ as derived using the numerical SWT with $U/t=1$. We note that the prescription developed above allows the calculations to be performed cheaply even for the $M=7$ chain using exact diagonalization, with the entire gate-voltage sweep obtained in a few minutes on a standard desktop computer.

The $M=1$ case is equivalent to a two-lead version of the AIM, and is a special limit since both source and drain lead couple to the same molecule orbital ($r_s=r_d=1$). This is the only `molecule' to satisfy `proportionate coupling', which allows us to reduce the system exactly to a single-channel model. This is achieved by canonical transformation of the lead operators from source-drain basis to even-odd basis,
\begin{align}
    c_{e\sigma} &=\tfrac{1}{\tilde{V}}\left ( V_s c_{s\sigma} + V_d c_{d\sigma} \right ) \;,\nonumber\\
    c_{o\sigma} &=\tfrac{1}{\tilde{V}}\left (  V_d c_{s\sigma} - V_s c_{d\sigma} \right ) \;,\label{eq:SI_eo}
\end{align}
where $\tilde{V}^2=V_s^2+V_d^2$, such that the odd lead is decoupled. This exact decoupling on the level of the bare Hamiltonian is not possible for $M>1$. For $M=1$ however, the model reduces to the 1-channel AIM and the usual SWT results then pertain on mapping to an effective 1-channel Kondo (1CK) model involving only the even lead \cite{Hewson}:
\begin{align}
    \hat{H}_{1CK} &= \hat{H}_{\rm lead}^e + J_{ee} \hat{\vec{\mathbb{S}}}\cdot \hat{\vec{s}}_{ee} + W_{ee}\sum_{\sigma} c_{e\sigma}^{\dagger}c_{e\sigma}^{\phantom{\dagger}} \label{eq:SI_1ck} \\
    J_{ee} &=2\tilde{V}^2\left [\frac{1}{U+\epsilon+V_g} - \frac{1}{\epsilon+V_g} \right ] \;,\\
    W_{ee} &=-\tilde{V}^2\left [\frac{1}{U+\epsilon+V_g} + \frac{1}{\epsilon+V_g} \right ] \;.
\end{align}
On inverting the transformation Eq.~\ref{eq:SI_eo} and returning to the physical source-drain lead basis, the effective 1-channel Kondo model Eq.~\ref{eq:SI_1ck} becomes precisely our original generalized 2CK model, Eq.~\ref{eq:SI_2ck}. However, the underlying single-channel description for $M=1$ implies additional relations between the coupling parameters that are not present in general \cite{mitchell2017kondo}. For $M=1$ we have $j_{ss}=j_{dd}=j_{sd}=j_{ds}$ and likewise $w_{ss}=w_{dd}=w_{sd}=w_{ds}$ (where, as before, we have divided out the hybridization dependence $J_{\alpha\beta}=V_{\alpha}V_{\beta}j_{\alpha\beta}$ and $W_{\alpha\beta}=V_{\alpha}V_{\beta}w_{\alpha\beta}$). The important conclusion is that the source-drain mixing terms $j_{sd}$ and $w_{sd}$ are equal to the local terms when there is an underlying 1CK description. 
Although the single-channel limit is certainly not generic for multi-orbital nanostructures, $j_{sd}$ and $w_{sd}$ are typically finite: these are the terms that mediate a source-drain current through the molecular junction. By contrast, for the 2CK critical point we require $j_{sd}=w_{sd}=0$, which corresponds to a conductance node.

\begin{figure}[t!]
\includegraphics[width=0.9\linewidth]{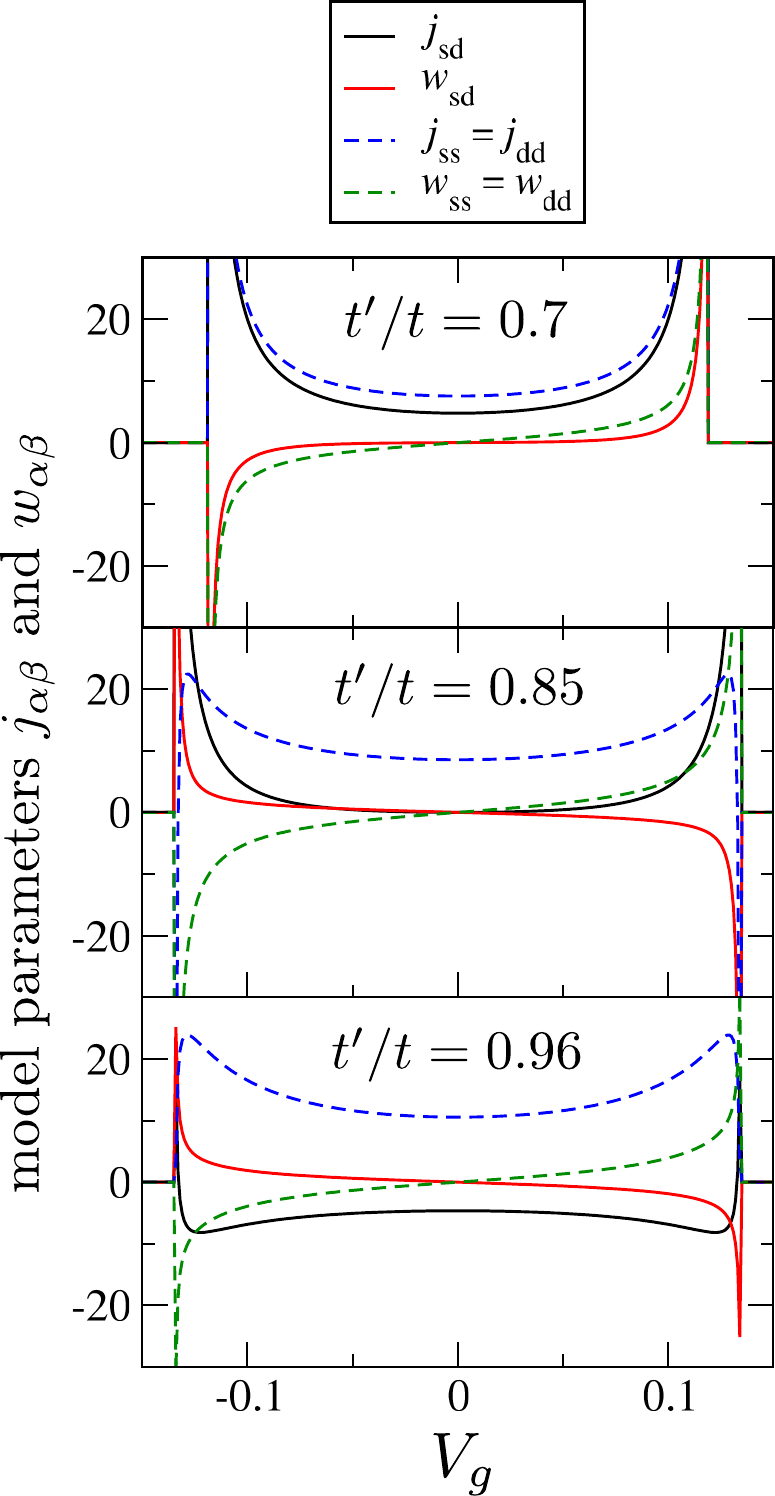}
  \caption{Effective model parameters $j_{\alpha\beta}$ and $w_{\alpha\beta}$ as a function of gate voltage $V_g$ from SWT calculations for the $M=5$ site molecular junction. We take local on-site interaction $U$ and nearest-neighbour hopping $t$, with $U/t=1$ and $t=0.5$ as usual. Here, the next-next-nearest-neighbour hopping $t'$  (connecting sites $1\leftrightarrow 4$ and sites $2\leftrightarrow 5$ of the chain) is now also included. The inclusion of $t'$ does not break exact particle-hole symmetry at $V_g=0$, but does continuously interpolate between $M=3$ and $M=5$ limiting cases of the pure chains shown in Fig.~\ref{fig:SI_chains} by providing a ``shortcut'' route through the molecule. This implies the existence of a node in $j_{sd}$ at $V_g=0$, which we confirm at a critical value of $t'=t'_c \simeq 0.4283$, as shown in the middle panel.
  }
  \label{fig:SI_swt_5site}
\end{figure}

Returning now to Fig.~\ref{fig:SI_chains}, we see that for $M=1$ the 1CK condition on the couplings is indeed satisfied. However, more interestingly, we find that $j_{ss}=j_{dd}=j_{sd}=j_{ds}$ and $w_{ss}=w_{dd}=w_{sd}=w_{ds}$ very accurately holds (although not exactly) for $M=5$, suggesting that (perhaps surprisingly) the odd lead combination is almost decoupled in the $M=5$ case at energies or temperatures $\ll U$ where the SWT effective model is expected to hold. This holds also for $M=9$ (not shown) and we conjecture is the case for all $M=4m+1$ (note however that the range of gate over which the doublet ground state occurs narrows on increasing $M$). Furthermore, for $M=3$ and $M=7$ (and again we suspect $M=4m+3$ in general), we have $j_{ss}=j_{dd} \simeq -j_{sd}=-j_{ds}$ and $w_{ss}=w_{dd} \simeq -w_{sd}=-w_{ds}$. This also implies an effective 1CK description at low energies or temperatures, but this time involving the \emph{odd} lead combination (and with the even lead decoupling).


\subsection{SWT results for 5-site cluster}\label{sec:si_5site}

The effective 1CK behavior of the odd-length Hubbard chains shown above is particular to the chain geometry around half-filling. Introducing couplings between different sites clearly produces greater complexity on the molecule. Here we consider the 5-site molecular cluster, Eq.~\ref{eq:SI_5site}, which has $sd$ symmetry, and also an exact $ph$ symmetry at $V_g=0$.

We expect interesting crossover behavior as a function of the ratio $t'/t$. For small $t'/t$, the $t'$ connections are a small perturbation and we expect physics similar to the pure 5-site chain. On the other hand, for large $t'/t$, we may regard $t$ as the perturbation . Source and drain leads then have a dominant connecting path that is of length 3; the $t'$ bonds offer a kind of `shortcut' through the $M=5$ chain. Tuning $t'/t$ therefore allows us to continuously deform between the $M=5$ and $M=3$ limits. Since the sign of $j_{sd}$ changes on going from $M=5$ to $M=3$ we expect a critical value $t'=t'_c$ where $j_{sd}=0$ just vanishes at $V_g=0$. As shown in Fig.~\ref{fig:SI_swt_5site}, this is precisely what we find. Here we plot the SWT model parameters for the model Eq.~\ref{eq:SI_5site}, with source and drain leads connected to sites 1 and 5, and for $U/t=1$ and $t=0.5$ as a function of gate voltage $V_g$. The ratio $t'/t$ is increased from top to middle to bottom panels. In the top panel, we see behavior similar to that of the pure $M=5$ chain shown in Fig.~\ref{fig:SI_chains}, but with the effective single-channel condition now violated ($j_{ss}=j_{dd} \ne j_{sd}$ and $w_{ss}=w_{dd}\ne w_{sd}$). In the bottom panel we see behavior instead more similar to the pure $M=3$ chain, with negative $j_{sd}$. The middle panel shows the `sweet spot' condition where $j_{sd}$ precisely vanishes at $V_g=0$ due to many-body QI effects of the competing pathways through the 5-site molecular moiety.


\section{QI Classes}\label{sec:jens}
Ref.~\cite{pedersen2014quantum} made a detailed study of many-body QI effects in molecular junctions, focusing on the transport problem within perturbation theory. 
Alternant hydrocarbon molecules were studied, whose structures can be accommodated on a bipartite graph with no odd loops, and to which the Coulson-Rushbrooke pairing theorem applies \cite{coulson1940note,tsuji2018quantum}.
Exchange cotunneling and potential scattering terms equivalent to our $J_{sd}$ and $W_{sd}$ were identified for systems with a net spin-doublet ground state.

A key insight from Ref.~\cite{pedersen2014quantum} is that QI effects in such systems can be classified according to the number of nodes found by varying the gate voltage all the way across a given charge state. In our notation, for a particular molecular junction they define,
\begin{equation}
    \mathcal{Q}_W = \frac{\langle \psi_0^{N;\tfrac{1}{2}}| d_{r_s\uparrow}^{\dagger} | \psi_0^{N-1;0}\rangle \langle \psi_0^{N-1;0} | d_{r_d\uparrow} |  \psi_0^{N;\tfrac{1}{2}} \rangle} {\langle \psi_0^{N;\tfrac{1}{2}}| d_{r_d\uparrow} | \psi_0^{N+1;1}\rangle \langle \psi_0^{N+1;1} | d_{r_s\uparrow}^{\dagger} | \psi_0^{N;\tfrac{1}{2}} \rangle} \;,
\end{equation}
and
\begin{equation}
    \mathcal{Q}_J = \frac{\langle \psi_0^{N;-\tfrac{1}{2}}| d_{r_s\downarrow}^{\dagger} | \psi_0^{N-1;0}\rangle \langle \psi_0^{N-1;0} | d_{r_d\uparrow} |  \psi_0^{N;\tfrac{1}{2}} \rangle} {\langle \psi_0^{N;-\tfrac{1}{2}}| d_{r_d\uparrow} | \psi_0^{N+1;0}\rangle \langle \psi_0^{N+1;0} | d_{r_s\downarrow}^{\dagger} | \psi_0^{N;\tfrac{1}{2}} \rangle} \;,
\end{equation}
where a sum over possible degeneracies of the $N\pm 1$ ground states is implied.

Ref.~\cite{pedersen2014quantum} showed that for bipartite molecules $\mathcal{Q}>0$ describes the \emph{odd} QI class with an odd number of QI nodes found upon scanning $V_g$, whereas $\mathcal{Q}<0$ describes the \emph{even} QI class with an even number of QI nodes. Indeed, $\mathcal{Q}_W>0$ if the source and drain leads connect to the same sublattice, whereas $\mathcal{Q}_W<0$ if the source and drain leads connect to different sublattices. Furthermore, bipartite molecules with an odd number of sites (which therefore have a spin-doublet ground state at their $ph$ symmetric point) satisfy the rule $\mathcal{Q}_W=-\mathcal{Q}_J$, meaning that $W$ and $J$ always necessarily belong to different QI classes. In a given molecule charge sector, if $W$ shows an odd number of nodes on scanning $V_g$ then $J$ must show an even number. This is borne out by our SWT results in Fig.~\ref{fig:SI_chains}, where the odd Hubbard chains (which have leads coupled to the same sublattice) have a single node in $W$ but no nodes in $J$. Ref.~\cite{pedersen2014quantum} concludes that a true conductance node is therefore `generally not possible' in such systems since $J_{sd}$ and $W_{sd}$ do not generally have a node at the same $V_g$.

However, it is also possible that two QI nodes in the even QI class can occur at the same gate voltage (degenerate case). This is precisely what we see in Fig.~\ref{fig:SI_swt_5site}. $W$ is in the odd QI class with a node in $W_{sd}$ at $V_g=0$. On the other hand $J$ is in the even QI class and has no nodes for $t'<t'_c$ but two nodes for $t'>t_c$. The two nodes coalesce and are exactly degenerate at $V_g=0$ when precisely at $t'=t'_c$. This is our 2CK critical point. The QI classification therefore provides a topological perspective on the 2CK phase transition in terms of coalescing and annihilating QI nodes in the 2CK model parameters.

Finally, we note that this analysis holds only for bipartite molecules. Therefore more general structures with odd loops (such as the one considered in Fig.~5 of the main text) need not be so constrained.


\section{Numerical Renormalization Group}\label{sec:SI_NRG}
As noted above, the SWT is a perturbative technique and more sophisticated methods are required to perform the exact mapping \cite{rigo2020machine}. In particular, the SWT does not take into account renormalization effects from the conduction electrons in the leads \cite{haldane1978theory}. It is also confined to the perturbative regime of large $U/V$. And of course, the \textit{solution} of the effective model is still required to understand the low-energy physics of the system and compute physical observables of interest relevant to experiments, such as the series conductance. 

The numerical renormalization group \cite{bulla2008numerical} (NRG) is considered the gold-standard method of choice for solving generalized quantum impurity problems, such as those describing the low-energy physics of molecular junctions \cite{mitchell2017kondo}. Furthermore, experimentally-relevant transport signatures can be accurately calculated using NRG down to low temperatures \cite{minarelli2022linear}, including renormalization effects from strong electron correlations, such as Coulomb blockade, Kondo effect, and many-body quantum interference phenomena. 
NRG is a numerically-exact, non-perturbative technique for solving quantum impurity type problems. The NRG method involves the following steps \cite{bulla2008numerical}.\\ 

(i) The source and drain lead density of states $\rho(\omega)$ is divided up into intervals on a logarithmic grid according to the discretization points $\pm D \Lambda^{-n}$, where $D$ is the bare conduction electron bandwidth, $\Lambda>1$ is the NRG discretization parameter, and $n=0,1,2,3,...$. The continuous electronic density in each interval is replaced by a single pole at the average position with the same total weight, yielding $\rho^{\rm disc}(\omega)$. In this work, as is conventional, we assume metallic leads with a constant density of states $\rho_0=1/2D$ within a flat band of half-width $D$.\\
(ii) The conduction electron part of the Hamiltonian $H_{\rm leads}$ is then mapped onto semi-infinite tight-binding chains for each lead $\alpha=s,d$:
\begin{eqnarray}
    H_{\rm leads} \to H_{\rm leads}^{\rm disc} = \sum_{\alpha,\sigma}\sum_{n=0}^{\infty} &t_n^{\phantom{\dagger}} \left ( f_{\alpha n,\sigma}^{\dagger}f_{\alpha n+1,\sigma}^{\phantom{\dagger}}+ {\rm H.c.} \right ) \;,\nonumber
\end{eqnarray}
where the Wilson chain coefficients $\{t_n\}$ are determined such that the density of states at the end of each chain reproduces exactly the discretized lead density of states, that is $-\tfrac{1}{\pi}{\rm Im}~\langle\langle f_{\alpha 0,\sigma}^{\phantom{\dagger}} ; f_{\alpha 0,\sigma}^{\dagger}\rangle\rangle = \rho^{\rm disc}(\omega)$. Due to the logarithmic discretization, the Wilson chain parameters decay roughly exponentially down the chain, $t_n \sim \Lambda^{-n/2}$. However the detailed form of the $t_n$ encode the specific lead density of states \cite{bulla2008numerical}.\\ 
(iii) The hybridizing molecule orbitals are coupled to site $n=0$ of the two Wilson chains. We define a sequence of Hamiltonians $H_N$ comprising the molecule and the first $N+1$ Wilson chain sites,
\begin{equation}\label{eq:Hn}
\begin{split}
    H_N=&H_{\rm mol} + H_{\rm hyb} + \sum_{\alpha,\sigma}\sum_{n=0}^{N-1}  t_n^{\phantom{\dagger}} \left ( f_{\alpha n,\sigma}^{\dagger}f_{\alpha n+1,\sigma}^{\phantom{\dagger}}+ {\rm H.c.} \right )  \;, \nonumber
\end{split}
\end{equation}
We now define the recursion relation,
\begin{equation}\label{eq:recursion}
    H_{N+1}=H_{N}+\sum_{\alpha ,\sigma}t_N^{\phantom{\dagger}} \left ( f_{\alpha N,\sigma}^{\dagger}f_{\alpha N+1,\sigma}^{\phantom{\dagger}}+{\rm H.c.} \right ) \;, \nonumber
\end{equation}
such that the full (discretized) model is obtained as $H^{\rm disc}=\lim_{N\to \infty} H_N$. \\
(iv) Starting from the molecule, the chain is built up successively by adding Wilson chain sites using the recursion, Eq.~\ref{eq:recursion}. At each step $N$, the intermediate Hamiltonian $H_N$ is diagonalized, and only the lowest $N_s$ states are retained to construct the Hamiltonian $H_{N+1}$ at the next step, with the higher energy states being discarded. With each iteration we therefore focus on progressively lower energy scales. Furthermore, the iterative diagonalization and truncation procedure can be viewed as an RG transformation \cite{bulla2008numerical}, $H_{N+1}=\mathcal{R}[H_N]$.\\
(v) The sequence of $H_N$ can be viewed as coarse-grained versions of the full model, which faithfully capture the physics at progressively lower and lower temperatures. Useful information is therefore extracted from each step, and physical observables can be obtained at essentially any temperature or energy scale \cite{bulla2008numerical}.

In this work we use a discretization parameter $\Lambda=3$ and retain $N_s=5000$ states at each NRG step. Total charge and spin projection abelian quantum numbers are exploited in the block diagonalization procedure. 

In the main text we present results for the linear response series differential conductance $G_c=dI_{sd}/dV_b$ through the molecular junction due to a small source-drain bias voltage $V_b$. Since anything more complex than the $M=1$ single-impurity limit does not satisfy the proportionate coupling condition, transport coefficients cannot be simply related to molecule Green's functions. However, the Kubo formula can still be used \cite{kubo1957statistical}. Here we use the `improved Kubo' formulation given in Ref.~\cite{minarelli2022linear}, which produces much more accurate results in the context of NRG calculations for the linear response electrical conductance.


\section{2CK critical point}\label{sec:SI_2CK}

\begin{figure}[t!]
\includegraphics[width=0.95\linewidth]{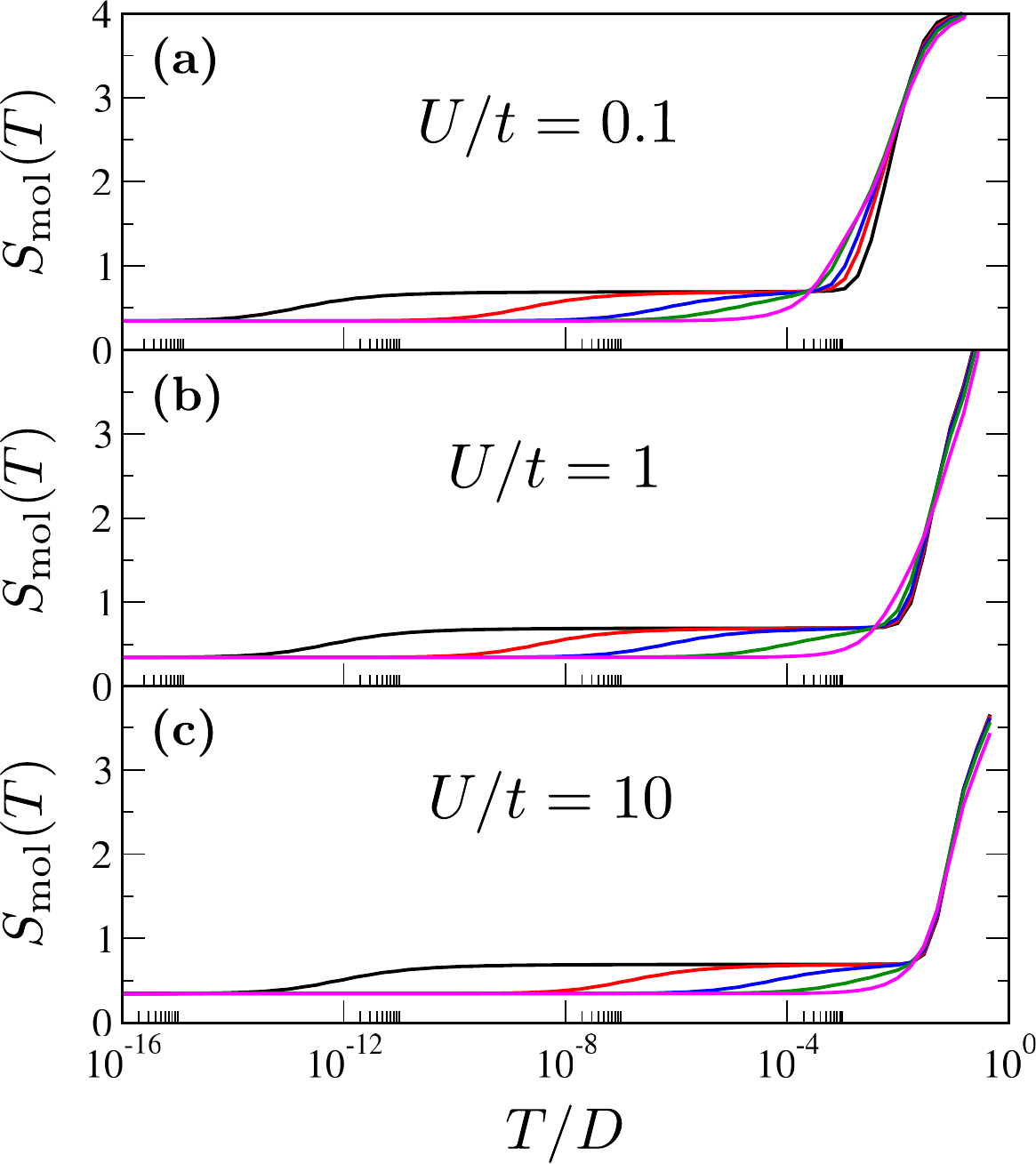}
  \caption{NRG results for the molecule contribution to the entropy $S_{\rm mol}$ as a function of temperature $T$ for the 5-site system tuned to the critical point at $t'=t'_c$, for different interaction strengths $U/t$ in panels (a), (b), and (c). As usual we set $t=0.5$ and the conduction electron bandwidth is $D=1$. In each case, the molecule-lead hybridization $V$ is varied, showing that the many-body QI-driven 2CK critical point with finite residual ($T=0$) entropy $S_{\rm mol}=\tfrac{1}{2}\ln(2)$ can be realized in all cases. The 2CK Kondo temperature $T_{\rm K}$ is seen to depend sensitively on $V$ (see Fig.~2 of the main paper); but we note that a remarkably large $T_{\rm K}\sim 10^{-2}U$ can be achieved in all cases, where no incipient local moment regime is observed and RG flow proceeds directly to the 2CK fixed point. For the black, red, blue, green and magenta lines we have $V=0.03, 0.04, 0.05, 0.06, 0.08$ for $U/t=0.1$ in panel (a); $V=0.1,0.125,0.15,0.2,0.3$ for $U/t=1$ in panel (b); and $V=0.3,0.4,0.5,0.6,0.75$ for $U/t=10$ in panel (c). The magenta lines show the maximum possible $T_{\rm K}$ achievable for each ratio of $U/t$.
  }
  \label{fig:SI_2ck_V}
\end{figure}

By tuning the ratio $t'/t$ in our 5-site molecular moiety Eq.~\ref{eq:SI_5site}, the SWT results indicate that we can simultaneously eliminate both $j_{sd}$ and $w_{sd}$ terms of the effective 2CK model. The $sd$ symmetry of the model means that the local Kondo couplings $j_{ss}=j_{dd}$ are equal. Since they are also antiferromagnetic, we expect to be able to realize the QI-2CK effect in this prototype molecular junction. In Fig.~\ref{fig:SI_2ck_V} we demonstrate that this is indeed the case, using full NRG results. Furthermore, the QI-2CK effect is found to be remarkably robust, holding beyond the perturbative regime of applicability of the SWT.

In Fig.~\ref{fig:SI_2ck_V} we plot the molecule contribution to the total entropy $S_{\rm mol}(T)$ as a function of temperature $T$ for systems tuned precisely to the QI-2CK critical point ($V_g=0$ and $t'=t'_c$). Panels (a,b,c) show different interaction strengths $U/t$ (with $t=0.5$ and $D=1$ fixed), over a very wide range from perturbative ($U/t=0.1$) to intermediate/non-perturbative strength ($U/t=1$), to strong coupling ($U/t=10$). We have also tested other ratios of $U/t$ and find equivalent results. The different lines plotted in each panel are for different molecule-lead hybridization strengths $V$. Small $V$ and large $U$ corresponds to the regime of applicability of the SWT calculation, but NRG allows to interrogate the non-perturbative regimes as well -- as shown in the figure. In particular, the magenta line in each panel is for a comparatively large $V$ which yields the largest 2CK Kondo temperature $T_{\rm K}$, while the black lines are for smaller $V$ where the corresponding $T_{\rm K}$ is then smaller.

Interestingly, we find that for all $U/t$ considered, we can access the QI-2CK effect for any $V$. The hallmark of the 2CK effect is the residual ($T=0$) molecule entropy of $\tfrac{1}{2}\ln(2)$ \cite{affleck1993exact}, which we here see in all cases. For smaller $V$ the Kondo scale $T_{\rm K}$ is small and we have a window over which the molecule entropy is $\ln(2)$. This corresponds to the local moment regime where the molecule hosts an essentially free spin-doublet state (this is the ground molecule doublet state that we project onto using the SWT). However, the most remarkable observation from our NRG results is that large $T_{\rm K}$ can be realized by tuning $V$. The largest $T_{\rm K}$ possible for each $U/t$ is in fact of order $10^{-2}U$, which is a high-energy scale when compared with the usual exponentially-small Kondo temperatures \cite{Hewson} found in regular single molecule junctions or semiconductor quantum dot devices. This is promising for experimental realization at accessible temperature scales. Finally, we note that as $U\to 0$, the maximum $T_{\rm K}$ also vanishes, meaning that as the non-interacting limit is approached, the 2CK effect also disappears. For typical interaction strengths in molecules or quantum dots, however, we expect the QI-2CK effect to be accessible at unusually high temperatures.

\begin{figure}[t!]
\includegraphics[width=0.85\linewidth]{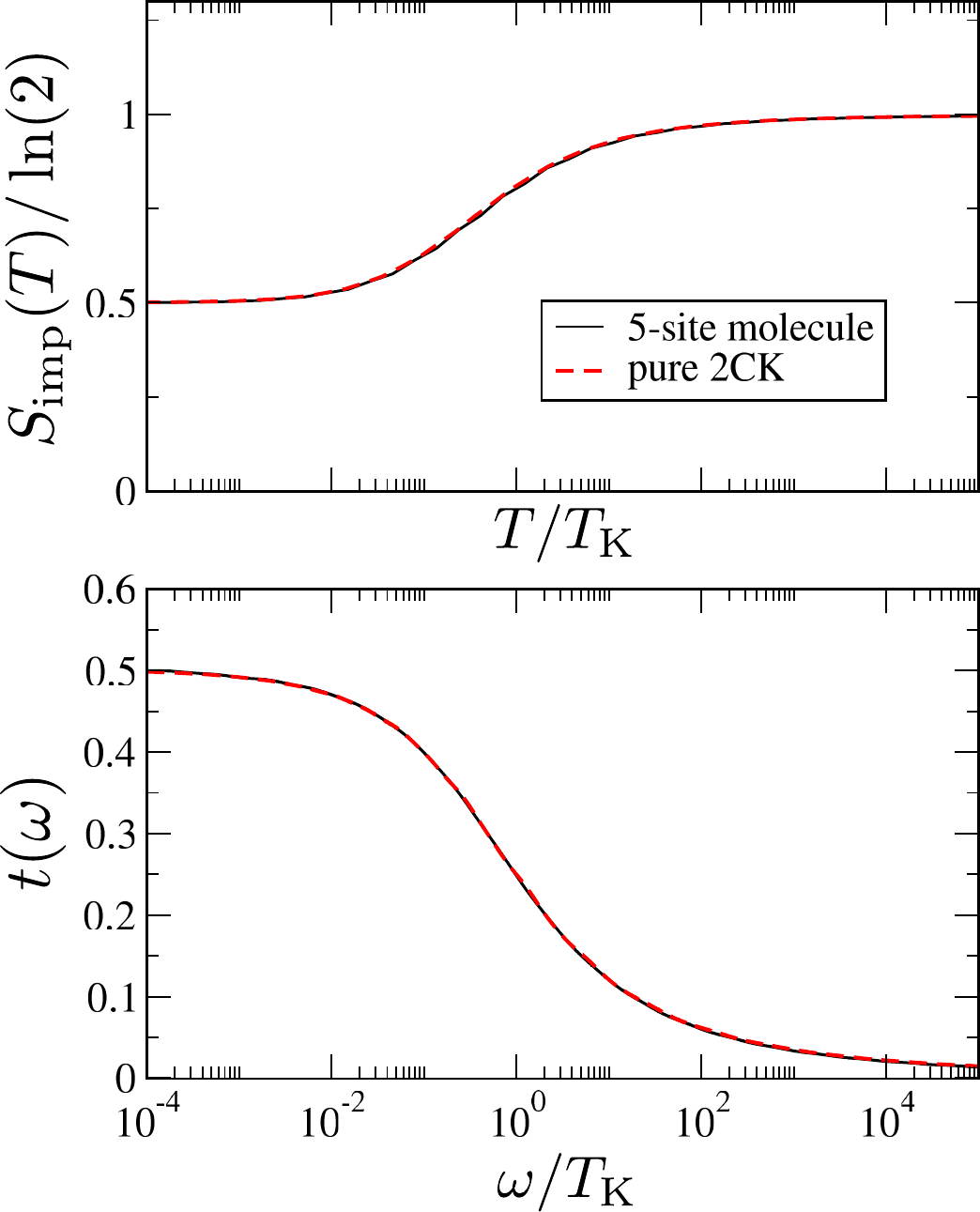}
  \caption{To confirm that the many-body QI-driven critical point in the 5-site molecular junction indeed realizes the classic 2CK critical point of the pure 2CK model, we show the `impurity' entropy $S_{\rm imp}(T)$ vs $T/T_{\rm K}$ (top panel) and the $T=0$ scattering t-matrix $t(\omega)$ vs $\omega/T_{\rm K}$ (bottom panel) rescaled in terms of the Kondo temperature $T_{\rm K}$ in the universal regime where $T_{\rm K}$ itself is $\ll U,D,t$. Full NRG results for the molecular junction (black solid lines) are compared with those for the pure 2CK model (dashed red lines), and agree perfectly. Shown for $U/t=10$, $t=0.5$, $D=1$, $V=0.3$, $V_g=0$, $t'=t'_c$.
  }
  \label{fig:SI_2ckuni}
\end{figure}

In Fig.~\ref{fig:SI_2ckuni} we confirm the SWT prediction that at the critical point of our 5-site molecule we realize the standard 2CK effect. For this, we select a molecule realization in the universal regime with a small $T_{\rm K}$. When scaled in terms of $T_{\rm K}$, physical observables should collapse to a universal curve characteristic of the pure 2CK model. The top panel of Fig.~\ref{fig:SI_2ckuni} shows NRG results for the entropy, rescaled in terms of $T/T_{\rm K}$ (black solid line), compared with standard results for the pure 2CK model. In the lower panel we show the energy-resolved conduction electron scattering t-matrix $t(\omega)$ at $T=0$ vs $\omega/T_{\rm K}$ for the molecule, compared with universal results of the pure 2CK model. Both agree perfectly in the universal regime considered. On the other hand, we note that the standard 2CK model presupposes the existence of an `impurity' spin-$\tfrac{1}{2}$ degree of freedom. Our results show that for larger $V$, the molecule hosts a 2CK critical regime that is not contained within the pure 2CK model since there is no incipient local moment regime. 


\section{FL crossover}\label{sec:SI_FL}

\begin{figure}[t!]
\includegraphics[width=\linewidth]{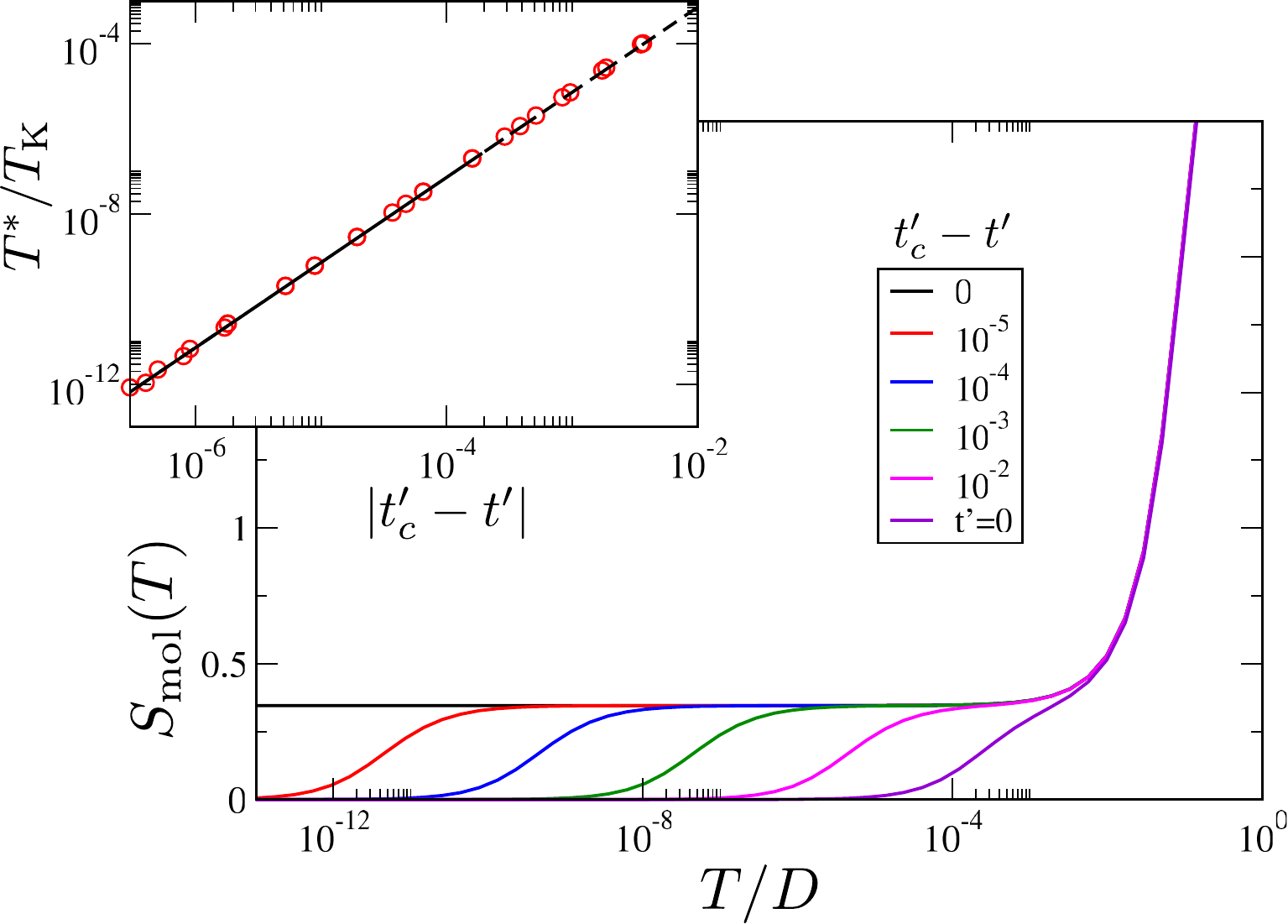}
  \caption{NRG results for the Fermi liquid crossover in the 5-site molecular junction near the QI-2CK critical point, with $t'$ successively approaching the critical point at $t'_c$. Main panel shows the entropy crossover, in which the critical $\tfrac{1}{2}\ln(2)$ entropy is quenched below $T^*$. Inset shows the dependence of $T^*$ itself on the perturbation $|t'_c-t'|$. Dashed line fit shows the robust power-law behavior $T^* \sim |t'_c - t'|^2$.
  We take $U/t=10$, $V=0.75$, $t=0.5$, $D=1$ and $V_g=0$.
  }
  \label{fig:SI_FL_tp}
\end{figure}

\begin{figure}[t!]
\includegraphics[width=\linewidth]{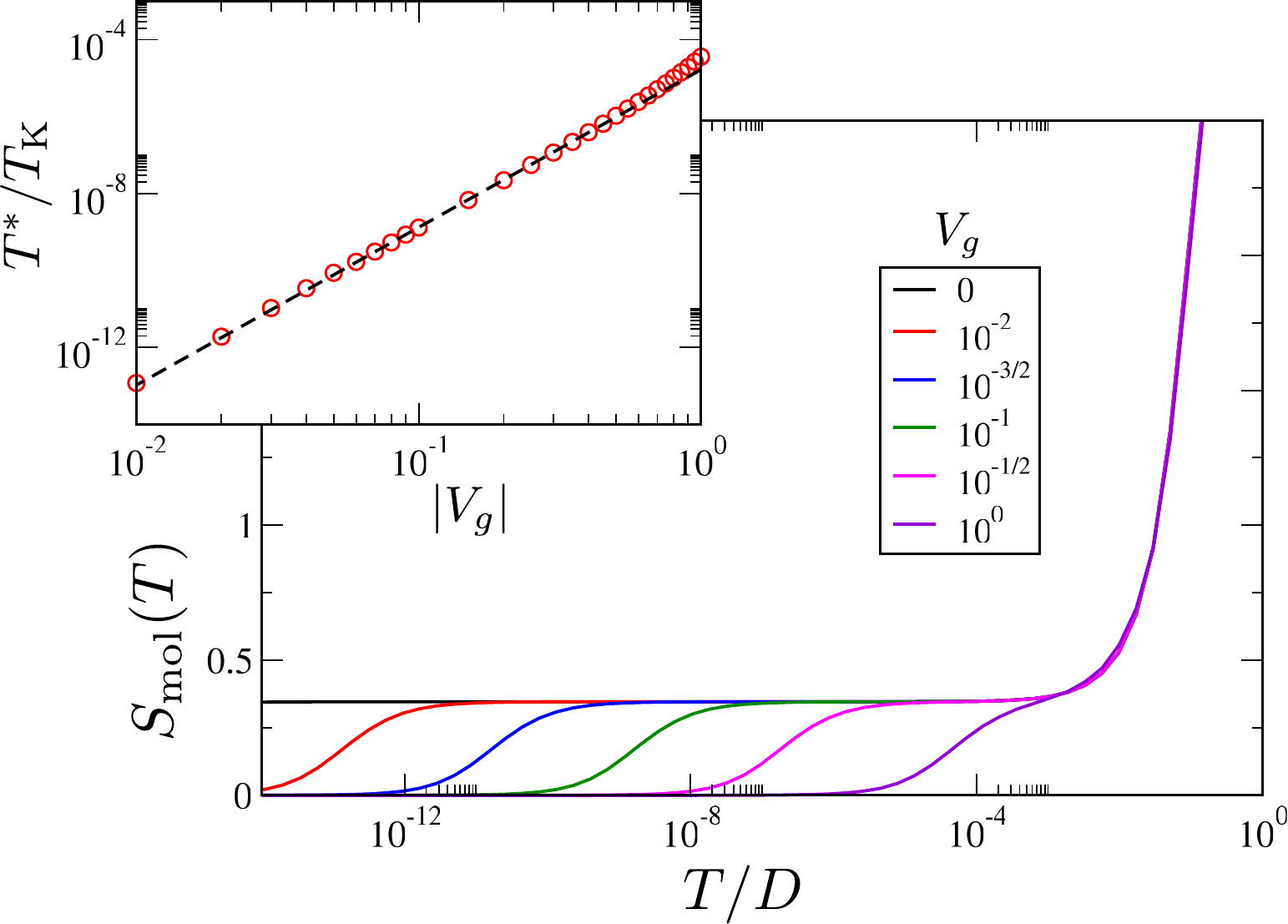}
  \caption{NRG results for the Fermi liquid crossover in the 5-site molecular junction, with $t'=t'_c$ but varying the gate voltage $V_g$ to destabilize the QI-driven 2CK critical point. Main panel shows the temperature-dependence of the entropy for each gate voltage setting. Inset shows the resulting crossover scale $T^*$ as a function of $V_g$, demonstrating with the black dashed line the power-law behavior $T^* \sim |V_g|^4$. We take $U/t=10$, $V=0.75$, $t=0.5$, $D=1$.
  }
  \label{fig:SI_FL_Vg}
\end{figure}

Having established that the 2CK critical point is accessible in molecular junctions by exploiting the many-body QI effect, we now wish to study in more detail the vicinity of the critical point and the relevant perturbations that drive the system along the NFL-FL crossover towards the stable fully-Kondo-screened ground state. We consider here two perturbations that destabilize the 2CK critical point in our 5-site molecular junction: deviations in the couplings away from $t'=t'_c$ (see Fig.~\ref{fig:SI_FL_tp}) and deviations in the gate voltage away from particle hole symmetry at $V_g=0$ (see Fig.~\ref{fig:SI_FL_Vg}).

When the perturbations are sufficiently small, the RG flow for the system proceeds first to the 2CK critical point (with characteristic $\tfrac{1}{2}\ln(2)$ entropy) on the scale of $T_{\rm K}$, and then it flows away from the critical fixed point and towards the FL ground state (with zero entropy) on the emergent low-energy scale $T^*$ \cite{affleck1993exact,Sela_2011,Mitchell_2012a}. For good scale separation $T^*\ll T_{\rm K}$ we expect a simple power-law dependence of the $T^*$ scale on the strength of the destabilizing perturbation. The power-laws reflect the scaling dimension of the leading RG relevant operators at the fixed point induced by the perturbation, and are a characteristic fingerprint of the critical fixed point. In the present context of the generalized 2CK model, we expect \cite{affleck1993exact} $T^* \sim a J_{sd}^2 + bW_{sd}^2$. But $J_{sd}$ and $W_{sd}$ themselves depend on $t'$ and $V_g$. In Fig.~\ref{fig:SI_FL_tp} we show that finite $|t'-t'_c|$ does indeed generate a finite FL scale $T^*$ and corresponding entropy crossover to a fully Kondo screened state. The inset shows specifically that
\begin{equation}
    T^* \sim |t'-t'_c|^2 \qquad :~ V_g=0
\end{equation}
Likewise in In Fig.~\ref{fig:SI_FL_Vg} we see that the FL crossover is also generated by finite $V_g$ even at $t'=t'_c$ with,
\begin{equation}
    T^* \sim |V_g|^4 \qquad ~~~~~:~ t'=t'_c
\end{equation}
The difference is because although $J_{sd}$ depends quadratically on both $t'$ and $V_g$, the critical value of $t'$ is at a finite value $t'_c$ while the critical value of $V_g$ is at $V_g=0$.

Finally, we comment on the robustness of the QI-2CK effect to changes in the molecule-lead couplings. The 2CK critical point is destabilized by anisotropy in the channel couplings, and we find,
\begin{equation}
    T^* \sim |V_c-V_L/V_R|^2 \;,
\end{equation}
where $V_c$ is the ratio $V_L/V_R$ at the critical point.


\section{Vibrations}\label{sec:SI_vib}
Vibrational effects can play a role in real molecular junctions \cite{yu2004inelastic} due to electron-phonon coupling. In fact, in some cases, molecular vibrations can quench single-particle QI \cite{ballmann2012experimental}. However, vibrational effects are expected to become less pronounced at low temperatures where 2CK criticality is predicted.

On a theoretical level, the effect of vibrations can be incorporated via a generalized Anderson-Holstein model \cite{hewson2001numerical}, which includes a coupling between the electrons and bosonic degrees of freedom. A modified version of the Schrieffer-Wolff transformation can still be devised in such a case to obtain an effective Kondo-type model, as worked out for the simplest molecular transistors in Ref.~\cite{paaske2005vibrational}. However, at weak electron-phonon coupling and low temperatures, we expect many-body QI to survive and the critical 2CK couplings to be simply renormalized. We leave a full treatment of such effects for future work.

Interestingly, Ref.~\cite{dias2009phonon} showed that electron-phonon coupling due to vibrations in a model of a C60 molecular junction can stabilize the 2CK effect.
Other reports on phonon-assisted 2CK effect can be found in e.g. Refs.~\cite{clougherty2003endohedral,lucignano2008two}.

Finally, we note that vibrational effects are far less important in artificial molecular junctions formed in semiconductor quantum dot devices.


\section{ML optimization and\\gradient descent}\label{sec:SI_GD}

Although we were able to identify conditions for the QI-2CK effect in molecular junctions in the preceding sections using a combination of symmetry arguments and an observation about the SWT for odd chains, a more systematic approach is required for inverse design in general. Indeed, even with the topology of the model Eq.~\ref{eq:SI_5site} set, some fine-tuning of the ratio $t'/t$ is still required. On the level of the SWT, one can fix all parameters except $t'$ and then use a simple root-finding algorithm to determine the value of $t'$ that yields $j_{sd}=0$. The critical value of $t'$ so determined will however only be an approximation to the true value of $t'_c$ found in full NRG calculations, since the SWT is of course only a perturbative approximation (albeit a rather good one). For the QI-2CK problem, one can again fix all parameters except $t'$ and run the NRG multiple times, varying $t'$, and search for the critical value $t'_c$ that minimizes the FL scale $T^*$. With only a single parameter to optimize, such a search is very fast and efficient, requiring  only a handful of `function evaluations' to converge exponentially quickly towards the critical point (here `function evaluation' means either doing the SWT mapping numerically to find $j_{sd}$ or performing an NRG run to determine the $T^*$ scale). Furthermore, the SWT result provides a good starting point guess for the NRG calculation.\\

However, inverse design of molecular junctions is in general much more complicated than the simplest constrained single-parameter optimization described above. We illustrate this in the main paper with the example of unconstrained optimization of the 4-site molecular cluster, demonstrating that machine learning (ML) methodologies can be applied to the numerical SWT calculation to realize the QI-2CK condition. We envision that this approach will be useful beyond the specific example of the 2CK problem, and will be applicable to the inverse design of molecular junctions for functionalities such as switching, rectification, thermoelectric properties etc, where the desired low-energy effective model parameters are known, but the optimal parent microscopic systems to realize them have yet to be identified.

\subsection{Inverse design using SWT}\label{sec:si_inv}
For molecular junctions, we imagine fixing the lead and hybridization Hamiltonians, and searching through the large configuration space of molecular structures that yield the desired effective model parameters from the SWT. For demonstration here we do not yet consider realistic molecules whose structure and chemistry are determined from \emph{ab initio} methods or from experimental analysis, but rather simply take the parameters of our generalized $\hat{H}_{\rm mol}$ molecular model to be freely tunable. Consider an abstract model for the isolated molecule of the form,
\begin{equation}
    \hat{H}_{\rm mol} = \sum_i \theta_i \hat{h}_i
\end{equation}
with coupling constants $\theta_i$ (which we assemble into a vector $\vec{\theta}$) and operators $\hat{h}_i$ that are defined on the molecule Hilbert space. Provided that the molecule ground state over the range of gate voltages of interest is a unique spin-double state, we can apply the numerical SWT prescription described above in Sec.~S-II. The desired effective model parameters are obtained by evaluating Eqs.~\ref{eq:SI_j}-\ref{eq:SI_A} and \ref{eq:SI_pM}-\ref{eq:SI_M}.

The desired functionality of the molecular junction places conditions on the effective model parameters $j_{\alpha\beta}$ and $w_{\alpha\beta}$. We encode these into a loss function $\mathcal{L}(\vec{\theta}) = f(j_{ss},j_{dd},j_{sd},w_{ss},w_{dd},w_{sd})$ where $f$ is some objective function to be minimized that defines the optimization landscape, whereas  $j_{\alpha\beta}$, $w_{\alpha\beta}$ depend implicitly on $\vec{\theta}$ through the SWT calculation. Despite the typically high complexity of $\hat{H}_{\rm mol}$, the ML optimization of the molecule can be done efficiently by gradient descent (GD). This involves calculating the derivative of the loss function $\vec{\nabla}_{\theta} \mathcal{L}$ with respect to the optimization parameters $\vec{\theta}$ and taking a finite step in the direction of the gradient vector from an initial guess $\vec{\theta}_0$ to a new guess $\vec{\theta}_1$ such that $\mathcal{L}(\vec{\theta}_1) < \mathcal{L}(\vec{\theta}_0)$. This is repeated iteratively until convergence within some tolerance to the minimum in $\mathcal{L}$, whence we have obtained the optimal set of coupling constants $\vec{\theta}$. 
We note however that care must be taken since the optimization landscape may contain many local minima or indeed degenerate global minima.

To make this more concrete, consider the generalized 4-site molecule model, Eq.~\ref{eq:si_4site}, with $U_m\equiv U=1$. We will treat the set of single-particle coupling constants $t_{mn}$ and the overall gate voltage $V_g$ as free tuning parameters that comprise the vector $\vec{\theta}$ to be optimized. Focusing on the QI-2CK effect, we demand a spin-doublet ground state giving an effective model for which $j_{sd}=0$ and $w_{sd}=0$, as well as $j_{ss}=j_{dd}>0$. A simple starting point for the loss function is,
\begin{equation}\label{eq:SI_loss}
    \mathcal{L}(\vec{\theta}) = a (j_{sd})^2 + b(w_{sd})^2 + c(j_{ss}-j_{dd})^2+d\sum_{\alpha}(|j_{\alpha\alpha}|-j_{\alpha\alpha})
\end{equation}
where $a,b,c,d$ are optimization hyperparameters that affect the optimization landscape and hence the learning curve. We also add a term that punishes gate voltages close to the Coulomb blockade steps, since the SWT breaks down in these regimes and the effective parameters $j_{\alpha\beta}$ and $w_{\alpha\beta}$ spuriously diverge. In practice, in a given iteration of the optimization procedure we fix a set of $t_{mn}$ and scan over the entire gate voltage range to identify regions with a spin-doublet ground state, and find the minimum of $\mathcal{L}$ for that gate sweep. Designing a good loss function is the art behind any complex optimization procedure.

In order to perform the GD, we would like the analytic gradient of the loss function. A numerical evaluation is costly when the number of parameters to be optimized is large, and finite-difference methods are inaccurate in practice. Although the analytic gradient with respect to the effective model parameters $j_{\alpha\beta}$ and $w_{\alpha\beta}$ is trivial, the desired gradient with respect to the \emph{bare} model parameters $\vec{\theta}$ is much more challenging. The key step is to work out the derivative of the particle and hole tunneling amplitudes in Eqs.~\ref{eq:SI_pM}, \ref{eq:SI_hM} with respect to a given tuning parameter $\theta_i$. 

An important implication of our SWT formulation is that the only non-trivial information needed to compute the gradient of the loss function analytically is the derivative of the ground  eigenstate $\partial_{\theta_i}\vec{U}_{\rm gs}$ and the derivative of the ground state energy $\partial_{\theta_i} E_{\rm gs}$. Suppressing indices to simplify the notation in the following, from Eq.~\ref{eq:SI_pM},
\begin{equation}
\begin{split}
    \partial_{\theta_i} p = &\left (  \partial_{\theta_i} \vec{U}_{\rm gs} \right )^{\dagger}\mathbb{M}_+ \vec{U}_{\rm gs}+  \vec{U}_{\rm gs}^{\dagger}\mathbb{M}_+\left ( \partial_{\theta_i} \vec{U}_{\rm gs}\right ) \\&+   \vec{U}_{\rm gs}^{\dagger}\Big (\partial_{\theta_i}\mathbb{M}_+ \Big ) \vec{U}_{\rm gs} 
    \end{split}
\end{equation}
and similarly for $\partial_{\theta_i} h$. Then from Eq.~\ref{eq:SI_M} it follows,
\begin{equation}
\begin{split}
    &\partial_{\theta_i} \mathbb{M}_{\pm} =\\ & \mathbb{X}_{\pm}^{\dagger}  \left [ E_{\rm gs} \boldsymbol{I} - \boldsymbol{H}_{\phi} \right ]^{-1}\left [\boldsymbol{h}^i_{\phi} -\partial_{\theta_i}  E_{\rm gs} \boldsymbol{I} \right ] \left [ E_{\rm gs} \boldsymbol{I} - \boldsymbol{H}_{\phi} \right ]^{-1} \mathbb{X}_{\pm}
    \end{split}
\end{equation}
where $\boldsymbol{h}^i_{\phi}=\vec{\phi}^{\dagger} \hat{h}_i ~\vec{\phi}$ such that $\boldsymbol{H}_{\phi}=\sum_i \theta_i \boldsymbol{h}^i_{\phi}$. Note that $\mathbb{X}_{\pm}$, $\boldsymbol{H}_{\phi}$ and $\boldsymbol{h}^i_{\phi}$ are all evaluated in the \emph{product} basis and are therefore independent of the optimization procedure. Only $\partial_{\theta_i}\vec{U}_{\rm gs}$ and $\partial_{\theta_i} E_{\rm gs}$ must be recomputed at each new optimization step. For the latter, we may simply use the Hellmann–Feynman theorem,
\begin{equation}
    \partial_{\theta_i} E_{\rm gs} = \langle \psi_{\rm gs} | \hat{h}_i | \psi_{\rm gs}\rangle \;,
\end{equation}
whereas the former follows from a standard result in non-degenerate perturbation theory \cite{feynman1939forces},
\begin{equation}
    \partial_{\theta_i} \vec{U}_{\rm gs} = \vec{\phi}^{\:\dagger}_{\rm gs}\sum_{j \ne {\rm gs}} \frac{\langle \psi_j | \hat{h}_i | \psi_{\rm gs}\rangle }{E_{\rm gs} - E_j} |\psi_j\rangle \;,
\end{equation}
where the sum runs over states in the ground quantum number subspace ($N,\sigma$) of the isolated molecule only.

These expressions allow us to compute accurately and cheaply the analytic gradient of the effective model parameters from the SWT, and hence the analytic gradient of any optimization loss function that depends on them. In turn, we may perform GD optimization of our effective model efficiently.

In the main text we demonstrate this with the example of a 4-site molecular cluster given by Eq.~\ref{eq:si_4site} using the loss function Eq.~\ref{eq:SI_loss}. Starting from a random guess for the parameters $\vec{\theta}_0$, we rapidly converge to a solution satisfying the QI-2CK conditions, as shown in Fig.~5 of the main paper. The parameters used in Fig.~5 resulting from our optimization procedure are $t_{11} \simeq 0.697$,~~$t_{22} \simeq 0.642$,~~$t_{33} \simeq 0.480$,~~$t_{44} \simeq 0.253$,~~$V_g=0.42$,~~$t_{12} \simeq 0.102$,~~$t_{13} \simeq 0.053$,~~$t_{14} \simeq 0.1$,~~$t_{23}\simeq 0.109$,~~$t_{24}\simeq 0.046$,~~$t_{34}\simeq 0.169$.


\subsection{Inverse design using NRG}
A more sophisticated treatment is provided using NRG which is a non-perturbative method \cite{bulla2008numerical}, and does not rely on the faithful derivation of a low-energy effective model \cite{rigo2020machine}, but can work directly with the bare model (at least for modest molecular complexity \cite{mitchell2017kondo}). Inverse design of molecular junctions using NRG affords the possibility of engineering loss functions for the optimization that work directly with the physical observables of interest. 

For example, in the present case of the QI-2CK effect, the quantum critical point is associated with a vanishing $T^*$ scale. This scale can be read off from the temperature dependence of the entropy, as shown in the main paper and in Figs.~\ref{fig:SI_FL_tp}, \ref{fig:SI_FL_Vg}. Defining $T^*$ through the relation $S_{\rm mol}(T=T^*)=\ln(2)/4$ and a loss function $\mathcal{L}(\vec{\theta})=T^*$ allows us to use ML methodologies to optimize the model parameters $\vec{\theta}$ to minimize $T^*$ by calculation of the molecule entropy. This is the approach we adopt in this work to find the critical value of $t'$ in the 5-site system. However, alternative approaches can be used. Fig.~4 of the main paper shows that the QI-2CK critical point is also associated with a conductance node at $T=0$. Therefore an alternative loss function might be $\mathcal{L}(\vec{\theta})=G_c(T=0)$.

A numerical (finite-difference) approximation to the gradient of the chosen loss function can be computed by running several standard NRG runs. NRG can then be integrated within a standard GD routine for inverse design model optimization. For constrained models with only a few optimization parameters, this might be sufficient.

However, for more complex problems (especially those for which each NRG run is computationally expensive and optimization efficiency is thus a priority), it is clearly advantageous to have the analytic gradient of the loss function from the NRG calculation. This is more challenging than for the SWT calculation because the NRG algorithm itself is much more complex. Also the loss function might comprise physical observables rather than simply effective model parameters. Fortunately this problem has been overcome recently by using \emph{differentiable programming} techniques \cite{bartholomew2000automatic} with the NRG method. This `automatic differentiable NRG' method allows exact gradients of NRG observables to be calculated quickly and efficiently \cite{rigo2022automatic}. This brings inverse design of molecular junctions using an NRG solver within reach.


\section{Non-interacting limit}\label{sec:SI_U0}

To better understand the QI-2CK critical point of the 5-site molecular junction studied above, here we consider the non-interacting $U\to 0$ limit of the model, 
\begin{equation}
\begin{split}\label{eq:SI_HU0}
    H_{\rm mol} =& 
    t\sum_{\sigma} \left( d_{1\sigma}^{\dagger} d_{2\sigma}^{\phantom{\dagger}}+d_{2\sigma}^{\dagger} d_{3\sigma}^{\phantom{\dagger}}+d_{3\sigma}^{\dagger} d_{4\sigma}^{\phantom{\dagger}}+d_{4\sigma}^{\dagger} d_{5\sigma}^{\phantom{\dagger}}+{\rm H.c.} \right ) \\
    &+ t'\sum_{\sigma}\left( d_{1\sigma}^{\dagger} d_{4\sigma}^{\phantom{\dagger}}+ d_{2\sigma}^{\dagger} d_{5\sigma}^{\phantom{\dagger}} +{\rm H.c.} \right ) \\
    H_{\rm hyb} =&  V\sum_{\sigma} \left ( c_{s\sigma}^{\dagger} d_{1\sigma}^{\phantom{\dagger}} + c_{d\sigma}^{\dagger} d_{5\sigma}^{\phantom{\dagger}} +{\rm H.c.} \right ) 
\end{split}
\end{equation}
which possesses $sd$-symmetry and particle-hole symmetry. Since $U=0$ the $\sigma=\uparrow$ and $\downarrow$ sectors are completely independent and can be treated separately. The two sectors are identical due to $SU(2)$ spin symmetry. Here we focus on the $\sigma=\uparrow$ sector for concreteness. 

\begin{figure}[t!]
\includegraphics[width=0.9\linewidth]{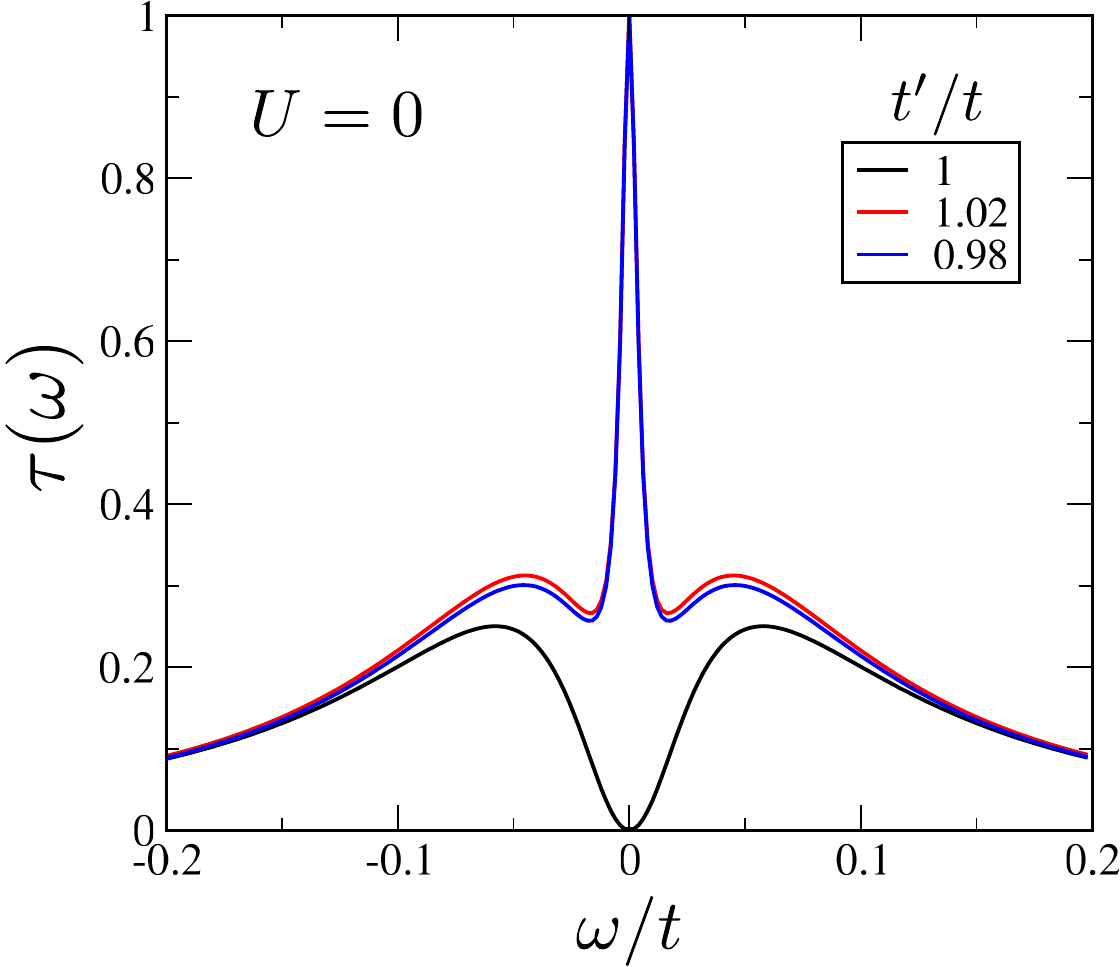}
  \caption{Transmission function $\tau(\omega)$ for the 5-site molecular junction in the non-interacting ($U=0$) limit, plotted for $t=1$ and $\Gamma_s=\Gamma_d=0.1$ for different $t'$. From Eq.~\ref{eq:SI_U0cond} the $T=0$ series conductance $G_c(0)=2e^2/h$ takes the maximum value for a single electron transistor for any $t'\ne t$, but precisely vanishes due to a single-particle QI node at $t'=t$. 
  }
  \label{fig:SI_U0}
\end{figure}

The first question is: what is the analogue of the QI-2CK critical point when $U=0$? The answer is provided by the SWT calculations shown in Fig.~2 of the main paper. For small but finite $U/t$ we see that the critical value of $t'$ tends to the value of $t$. This is supported by full NRG calculations. Thus, we deduce that $t'/t\to 1$ as $U/t \to 0$. Below, we therefore examine Eq.~\ref{eq:SI_HU0} in the vicinity of the point $t'/t=1$. Indeed, we find that $t'/t=1$ is a special point.

We start by performing a rotation to an even/odd orbital basis, viz:
\begin{eqnarray}
    d_{1 \substack{e\\o} \sigma}&=\tfrac{1}{\sqrt{2}}[d_{1\sigma} \pm\ d_{5\sigma}] \;\\
    d_{2 \substack{e\\o} \sigma}&=\tfrac{1}{\sqrt{2}}[d_{2\sigma} \pm\ d_{4\sigma}] \;\\
     c_{\substack{e\\o} \sigma}&=\tfrac{1}{\sqrt{2}}[c_{s\sigma} \pm\ c_{d\sigma}] \;.
\end{eqnarray}
In this basis, the model exactly decouples into even and odd sectors,
\begin{equation}
\begin{split}\label{eq:SI_HU0eo}
    H_{\rm mol} =& 
    \sum_{\sigma} \left[ (t+t')d_{1e\sigma}^{\dagger} d_{2e\sigma}^{\phantom{\dagger}}+\sqrt{2}t d_{2e\sigma}^{\dagger} d_{3\sigma}^{\phantom{\dagger}}+{\rm H.c.} \right ] \\
    &~~~+ (t-t')\sum_{\sigma}\left[ d_{1o\sigma}^{\dagger} d_{2o\sigma}^{\phantom{\dagger}}+ {\rm H.c.} \right ] \\
    H_{\rm hyb} =&  V\sum_{\sigma} \left ( c_{e\sigma}^{\dagger} d_{1e\sigma}^{\phantom{\dagger}} + c_{o\sigma}^{\dagger} d_{1o\sigma}^{\phantom{\dagger}} +{\rm H.c.} \right ) 
\end{split}
\end{equation}
Importantly, we see that the orbital $d_{2o\sigma}$ exactly decouples from the rest of the system when $t'=t$, and sits at zero energy. This directly implies that the $T=0$ contribution from the molecule to the total entropy of the system is $S_{\rm mol}(0)=\ln(4)$ at this special point, since all other molecule degrees of freedom are screened by the leads. 

By contrast, we know from our NRG results at small but finite $U/t$ where the critical point arises at $t'\simeq t$ that the $T=0$ residual molecule entropy is $\tfrac{1}{2}\ln(2)$. The critical point is therefore a genuine non-Fermi liquid (NFL) in the sense that it is not adiabatically connected to the non-interacting limit. However, the energy scale $T_{\rm K}$ corresponding to the formation of the NFL state also vanishes as $U/t \to 0$ as argued above. We speculate that it is the decoupled single-particle state $d_{2o\sigma}$ of the non-interacting molecule that seeds the NFL state when interactions are turned on.

Finally, we consider dynamical properties of the $U=0$ molecule. The molecule Green's functions of the lead-coupled system can be simply determined,
\begin{equation}\label{eq:SI_GF}
    \boldsymbol{G}_{\rm mol}^{U=0}(\omega) = \begin{bmatrix}
\omega^+ + i\Gamma & ~-t~ & 0 & ~-t'~ & 0\\
-t & \omega^+ & ~-t~ & 0 & -t' \\
0 & -t & \omega^+ & -t & 0 \\
-t' & 0 & -t & \omega^+ & -t \\
0 & -t' & 0 & -t & \omega^+ +i\Gamma
\end{bmatrix}^{-1}
\end{equation}
where $\left [\boldsymbol{G}_{\rm mol}\right ]_{ij}=\langle\langle d_{i\sigma}^{\phantom{\dagger}} ; d_{j\sigma}^{\dagger}\rangle\rangle$, $\omega^+=\omega+i0^+$ and $\Gamma=\pi V^2 \rho_0$, with $\rho_0$ the lead density of states (we have assumed the wide-flat-band limit for simplicity). 

Of particular interest is the through-molecule Green's function that connects the source and drain leads, $G_{15}(\omega)$, since this controls series transport in the non-interacting limit via the Landauer formula \cite{landauer1957spatial,minarelli2022linear},
\begin{equation}\label{eq:SI_U0cond}
    G_c(T) = \left(\frac{2e^2}{h}\right )\int d\omega \frac{-\partial f(\omega)}{\partial \omega} \times \tau(\omega) \;,
\end{equation}
where $G_c(T)$ is the series electrical conductance $dI_{sd}/dV_b$ resulting when a current $I_{sd}$ flows due to a bias voltage $V_b$ between source and drain leads in linear response, and $f(\omega)=1/[1+\exp(\omega/T)]$ is the equilibrium fermi function. The transmission function $\tau(\omega)$ is related to the molecule Green's functions via $\tau(\omega)=|2\Gamma G_{15}(\omega)|^2$. Therefore, the $T=0$ conductance  follows as $G_c(0)=(2e^2/h)\times |2\Gamma G_{15}(0)|^2$ and depends on the fermi level behavior of the through-molecule Green's function $G_{15}(0)$. Analysis of Eq.~\ref{eq:SI_GF} shows that,
\begin{equation}
    2\Gamma G_{15}^{U=0}(\omega\to 0) = \begin{cases}
-i & ~~:~t' \ne t\\
0 & ~~:~t'=t
\end{cases}
\end{equation}
The full energy-dependence of the transmission function is shown in Fig.~\ref{fig:SI_U0}, and establishes that for $t'\ne t$ a narrow peak pinned at the fermi energy to $\tau(0)=1$ collapses at $t'=t$. Thus the $T=0$ conductance $G_c(0)=2e^2/h$ for any $t'\ne t$ but has an exact conductance node $G_c(0)=0$ for $t'=t$. This is a result of the single-particle QI in the system at $t'=t$, giving rise to destructively interfering paths through the molecule. We note that in the interacting case $U>0$ the $T=0$ conductance also saturates to its maximum $G_c(0)=2e^2/h$ away from the QI-2CK critical point $t'\ne t'_c$, but series conductance vanishes $G_c(0)=0$ at the critical point $t'=t'_c$. However, in this case the conductance node is a true many-body QI effect.


\section{Candidate molecules for QI-2CK}
In the main text of the paper, the focus was on identifying the \textit{simplest} model systems to realize the QI-2CK effect. Leveraging parity (mirror) symmetry and particle-hole symmetry, we found a simple 5-site Anderson-type model with a single tuning parameter. Relaxing the symmetry constraints yielded a solution with only 4 sites, but at the expense of  fine-tuning of more parameters.

\begin{figure*}[t]
\includegraphics[width=\linewidth]{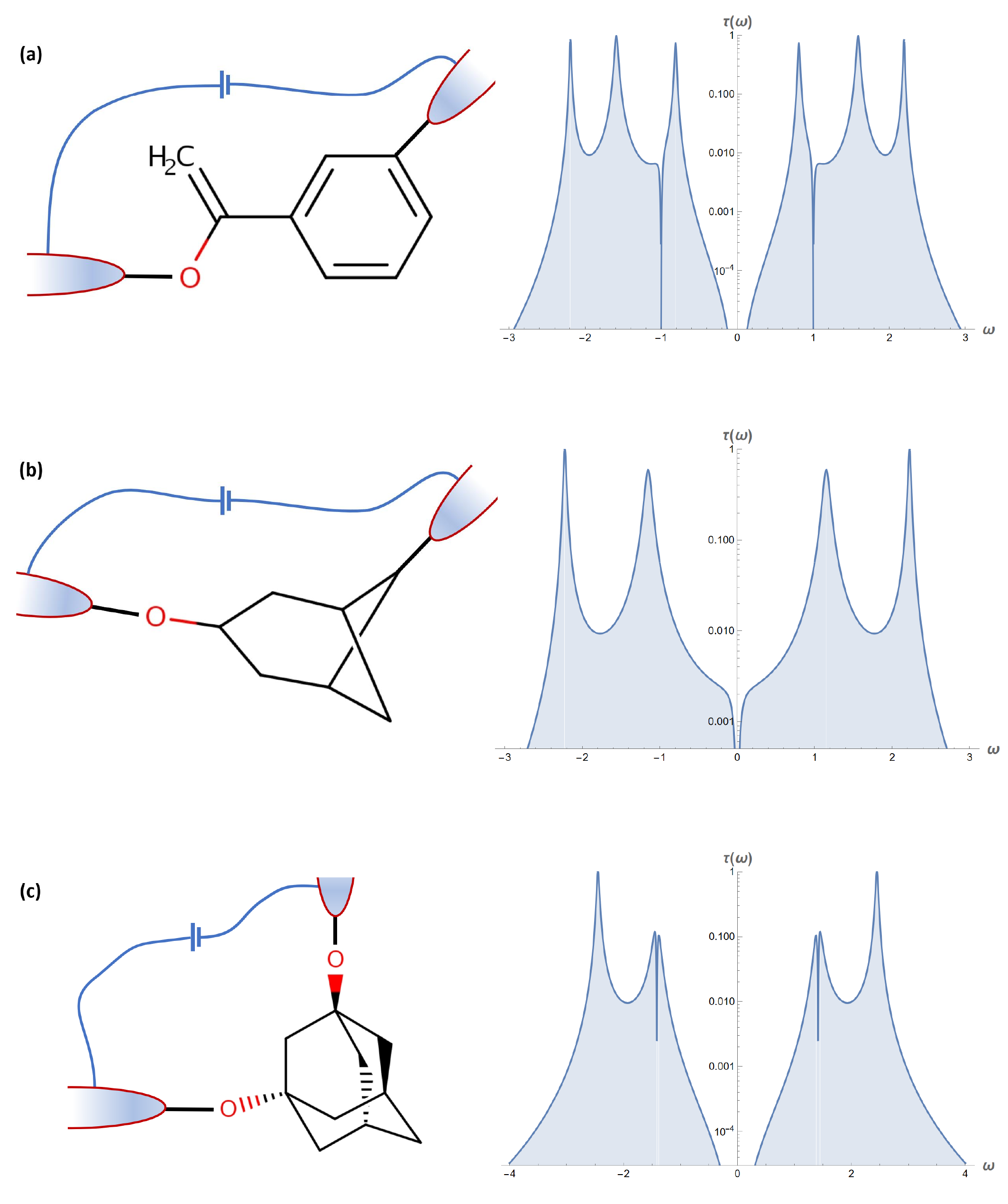}
  \caption{Non-interacting limit of candidate molecular junctions to realize the QI-2CK effect. The through-molecule transmission functions plotted on the right correspond in each case to the junction illustrated on the left. Parent molecule for each junction is (a) 1-phenylethen-1-ol; (b) bicyclo[3.1.1]heptan-3-ol; (c) adamantane-1,3-diol. See text for details.
  }
  \label{fig:SI_mol}
\end{figure*}

As emphasized in the main text, these toy models are not intended to be descriptions of any realistic molecule, but rather they are instructive examples of structural motifs containing the necessary orbital complexity for many-body QI. However, in the context of realizing nontrivial many-body physics in semiconductor quantum dot devices, we note that simpler structures are desirable, and fine-tuning can be performed \textit{in situ} \cite{barthelemy2013quantum}.

For single-molecule junctions, many-body methods \cite{pedersen2014quantum} can be interfaced with \textit{ab initio} techniques \cite{lucignano2009kondo} and the inverse design methodology presented here, to search for realistic molecules exhibiting the QI-2CK effect. This is beyond the scope of the present work, but here we propose some candidate molecules, based on simple design principles learned from our study of the toy models.

From the previous sections, we found that the following ingredients in a microscopic model will lead to many-body QI of the right type to produce 2CK physics: 
particle-hole symmetry, such that (i) $\epsilon_j=-U_j/2$ on all sites and (ii) no odd-membered loops in the tunnel-coupling geometry. This guarantees a half-filling condition and hence a net spin $S=1/2$ ground state, as well as $W_{sd}=0$. To get $J_{sd}=0$ as well, we need (iii) at least two distinct paths through the molecule connecting source and drain leads, with different odd numbers of sites (for example, one path with 3 sites between source and drain, and another with 5 sites); and (iv) a means of tuning the relative importance of these paths (for example, different hoppings $t$ along one path and $t'$ along another, or different onsite potentials etc). Mirror symmetry is not required if one can tune the molecule-lead couplings.

\begin{figure*}[t]
\includegraphics[width=\linewidth]{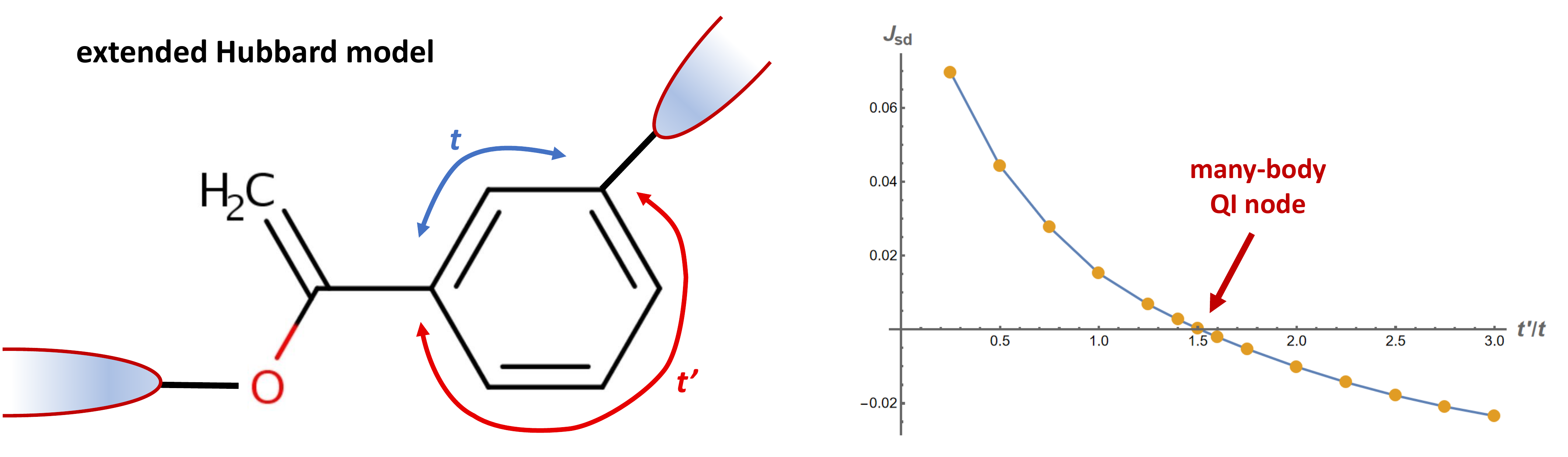}
  \caption{Hubbard model for candidate molecule (a), with asymmetric ring tunnelings $t$ and $t'$ as indicated (left). Numerical SWT calculation of $J_{sd}$ (right) shows a many-body QI node as a function of tuning $t'/t$, where $J_{sd}=0$. At this point the QI-2CK effect would be expected at low temperatures. See text for details.
  }
  \label{fig:SI_molSWT}
\end{figure*}

We also found that the non-interacting limit can provide useful insights. The QI-2CK effect in the many-body ($U>0$) case features a conductance node, which appears to be continuously connected to a conductance node due to single-particle QI in the non-interacting ($U=0$) limit. For $U>0$ we have a residual finite entropy at the 2CK critical point due to a decoupled Majorana degree of freedom; in the $U=0$ limit we also have a finite $T=0$ entropy, due to a decoupled effective fermionic degree of freedom. These features provide a good starting point guide to search for real molecules that might exhibit QI-2CK.

In Fig.~\ref{fig:SI_mol} we consider three such candidate molecular junctions, whose parent molecules are fairly standard and commercially available. We acknowledge but do not address here the practical challenges associated with joining the molecules to the leads in the correct geometry. We do not attempt a first-principles type study, but simply take a non-interacting tight-binding model for each, assuming equal nearest-neighbour tunneling matrix elements $t\equiv 1$. The leads are connected to the sites indicated, and for the purpose of demonstration we set $\Gamma_s=\Gamma_d=0.1t$. The proposed structures fulfil the above design criteria.

In Fig.~\ref{fig:SI_mol} panel (a) we consider a molecular junction whose parent molecule is 1-phenylethen-1-ol. Under the tight-binding assumption, the conductance follows from the transmission function $\tau(\omega)$ using Eq.~\ref{eq:SI_U0cond}. The transmission function is plotted on the right as a function of frequency $\omega$. As desired, we see that $\tau(0)=0$ and hence the conductance $G_c(0)=0$. At the single-particle level we also have a decoupled effective degree of freedom with zero energy eigenvalue.

In panel (b) we consider a junction with parent molecule bicyclo[3.1.1]heptan-3-ol, whereas in panel (c) we consider an adamantane-1,3-diol. 
Again in both cases, we have a vanishing transmission function at the Fermi level, and an effective decoupled degree of freedom.

Having identified plausible structures based on the non-interacting limit, we now take molecule (a) and formulate a simple Hubbard (PPP) type model. We do this by introducing a finite local $U=1$ on all sites (and therefore set $\epsilon=-0.5$ on all sites to ensure particle-hole symmetry). We find that the ground state is, as desired, a net spin $S=1/2$ doublet state, and perform numerically the SWT to obtain the generalized 2CK model, Eq.~2. As expected, $W_{sd}=0$ automatically by symmetry. However, even though the $U=0$ structure with equal couplings had a $T=0$ conductance node, for $U>0$ we have a finite $J_{sd}$ indicating finite conductance. Therefore the unperturbed structure does not realize 2CK. This is the same story as for the 5-site toy model structure in the main text: one has to fine-tune the many-body QI in the interacting case to achieve QI-2CK via the vanishing $J_{sd}=0$. We do that here by introducing different tunneling matrix elements $t$ and $t'$ around the benzene ring, as illustrated in Fig.~\ref{fig:SI_molSWT}. In practice, this perturbation might be achievable by differential gating or functionalization of certain sites around the ring. We now repeat the SWT and plot $J_{sd}$ as a function of $t'/t$, see right plot in Fig.~\ref{fig:SI_molSWT}. As hoped for, we do find a sweet spot where many-body QI effects produce an exact node in $J_{sd}$. At this point, the molecular junction will exhibit low-temperature 2CK physics.

Naturally, these calculations are simplistic, but they gesture towards the kind of simulations that would be possible for real molecules using \textit{ab initio} methods.


\section{Outlook: three-channel\\Kondo effect}
\begin{figure}[t]
\includegraphics[width=0.9\linewidth]{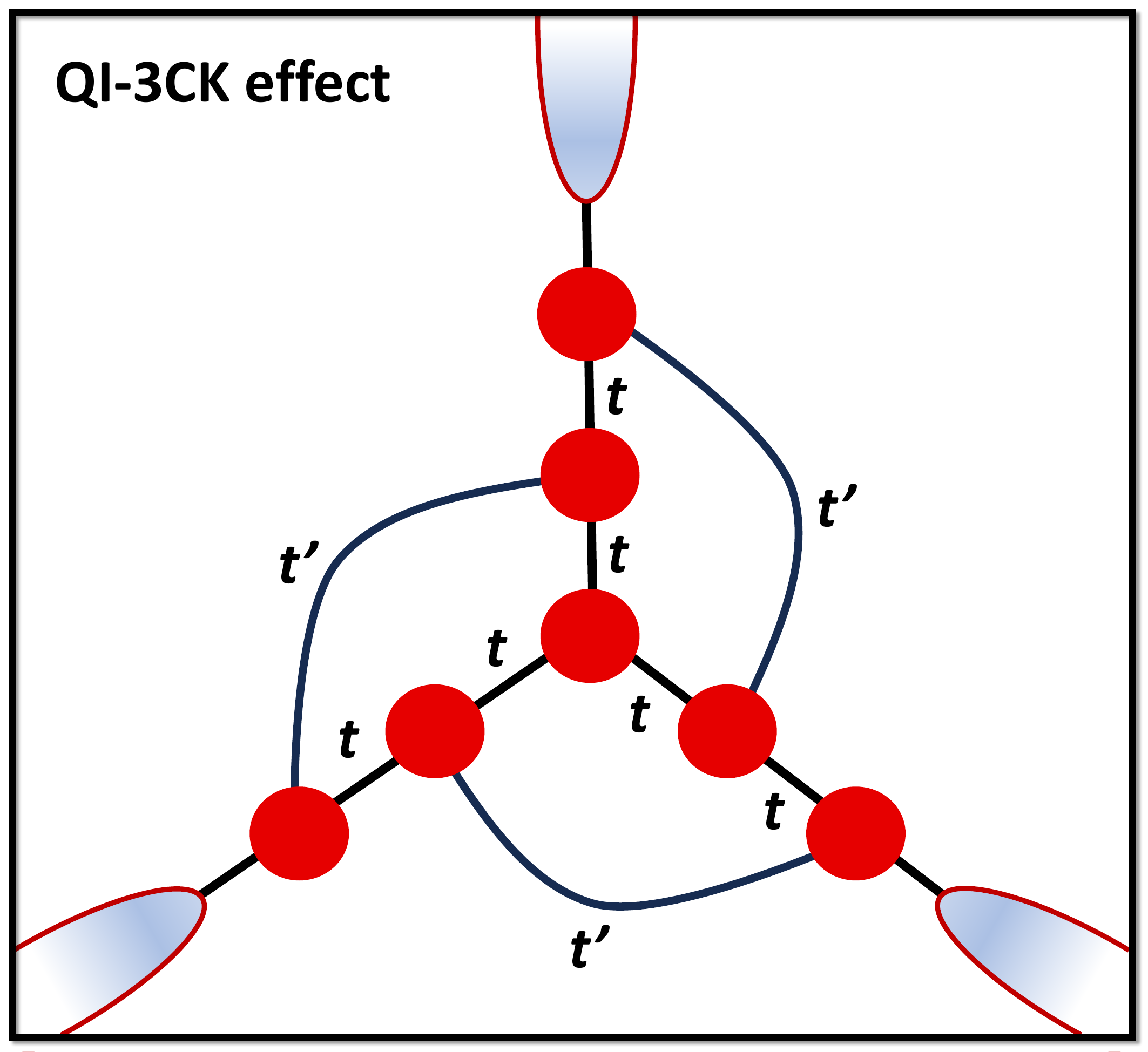}
  \caption{Proposed 7-site structure to realize the 3CK effect using many-body QI.
  }
  \label{fig:SI_3ck}
\end{figure}

Finally, we touch briefly on an outlook towards realizing other kinds of physics by exploiting many-body QI effects. One such simple extension is to the three-channel Kondo (3CK) model. This was realized in hybrid metal-semiconductor island nanostructure experiments in Ref.~\cite{iftikhar2018tunable}, but we note here that 3CK can also be found in simple few-orbital structures, precisely analogous to the 2CK effect found in Fig.~1 of the main text. The design principle is to find a structure with an odd number of interacting sites at particle-hole symmetry, with no odd loops, where there are two routes connecting each pair of leads -- one comprising 3 sites and the other with 5 sites. For the two-lead setup, Eq.~S-7 realizes the 2CK model in the simplest 5-site structure upon fine-tuning the ratio $t'/t$. Generalizing this to a three-lead setup, the simplest structure has 7 interacting sites. We have $\hat{H}=\hat{H}_{\rm mol}+\hat{H}_{\rm leads}+\hat{H}_{\rm hyb}$ where,
\begin{equation}
\begin{split}
    \hat{H}_{\rm mol}=&U\left( \hat{n}_{0\uparrow}-\tfrac{1}{2}\right)\left (\hat{n}_{0\downarrow}-\tfrac{1}{2}\right) \\&+ U\sum_{\alpha=1,2,3}\;\sum_{n=1,2}\left( \hat{n}_{\alpha n \uparrow}-\tfrac{1}{2}\right)\left (\hat{n}_{\alpha n\downarrow}-\tfrac{1}{2}\right)\\
&+t\sum_{\alpha,\sigma}\left(d_{0\sigma}^{\dagger}d_{\alpha 1 \sigma}^{\phantom{\dagger}} +d_{\alpha 1 \sigma}^{\dagger}d_{\alpha 2 \sigma}^{\phantom{\dagger}} + {\rm H.c.}\right) \\
&+t'\sum_{\sigma}\left(d_{12 \sigma}^{\dagger}d_{21 \sigma}^{\phantom{\dagger}}+d_{22 \sigma}^{\dagger}d_{31 \sigma}^{\phantom{\dagger}}+d_{32 \sigma}^{\dagger}d_{11 \sigma}^{\phantom{\dagger}} + {\rm H.c.}\right) 
\end{split}
\end{equation}
where $\alpha=1,2,3$ runs over the three `branches'. $\hat{H}_{\rm leads}$ is given by Eq.~S-3, generalized to three leads. Each lead is coupled to the end of each branch,
\begin{equation}
    \hat{H}_{\rm hyb}=V\sum_{\alpha,\sigma}\left ( d_{\alpha 2 \sigma}^{\dagger}c_{\alpha \sigma}^{\phantom{\dagger}} + {\rm H.c.}
\right ) \;.
\end{equation}
The setup is illustrated in Fig.~\ref{fig:SI_3ck}.

The SWT yields a model of the same form as Eq.~2 of the main text, except now $\alpha=1,2,3$ and $\beta=1,2,3$ run over the three channel indices. Note that by the three-fold symmetry of the structure we automatically have equal diagonal elements $J_{\alpha\alpha}\equiv J>0$. The QI-3CK effect arises when all off-diagonal $J_{\alpha\ne\beta}$ and $W_{\alpha\ne\beta}$ elements vanish. Due to the particle-hole symmetry inherent to this model, the latter are strictly zero. Therefore we just require many-body QI effects to kill off the $J_{\alpha\ne\beta}$ terms. The same arguments used for the QI-2CK effect show that this can be achieved for 3CK by again tuning the ratio $t'/t$.



%